\documentclass[notitlepage,showkeys]{revtex4-1}
\usepackage{color,bm}
\usepackage{graphicx}
\usepackage{amsmath,amssymb}

\begin{document}
\title{Subgap in the surface bound states spectrum of Superfluid $^3$He-B with Rough Surface}
\author{Y. Nagato} 
\email{nagato@riise.hiroshima-u.ac.jp}
\author{S. Higashitani}
\author{K. Nagai}
\affiliation{Graduate School of Integrated Arts and Sciences, Hiroshima University, Kagamiyama 1-7-1, Higashi-Hiroshima, 739-8521 Japan}


\begin{abstract}
Subgap structure in  the surface bound states spectrum of Superfluid
 $^3$He-B
with rough surface is discussed. The subgap is formed by the
level repulsion between the surface bound state and the continuum
states in the course of multiple scattering by the surface roughness. 
We show that the level repulsion is originated from the
nature of the wave function of the surface bound state
that is now recognized as Majorana Fermion. We study the superfluid
$^3$He-B with a rough surface and under magnetic field perpendicular
to the surface using the quasi-classical Green function together with
random S-matrix model. We calculate the self-consistent order parameters,
the spin polarization density and the surface density of states. It is shown that the subgap is
found also under the magnetic field perpendicular to the surface. 
Magnetic field dependence of the transverse acoustic impedance
is also discussed.
\end{abstract}


\maketitle

 \section{Introduction}
Since the discovery of superfluid ${}^3$He, 
it has been well known that the 
superfluid state of unconventional paring states 
is significantly affected by the presence of surface.\cite{AdR} 
The order parameter is  suppressed
near the boundary within some coherence lengths and the
surface density of states is considerably modified from the bulk 
behavior. Such surface effects are caused by the surface
scattering and depend upon the nature of boundary condition, 
whether quasi-particle scattering is specular or diffusive.

Typical example of the surface effects is the formation of
surface bound states with energy below the bulk energy gap.
\cite{BZ,HN,Zhang,Nagai2008}
Buchholtz and Zwicknagl\cite{BZ} found mid-gap states in the
BW state with a specular surface. Hara and Nagai\cite{HN}
showed that $p$-wave polar state with a specular surface  always 
has  zero energy surface 
bound states irrespective of the direction of the Fermi momentum. 
Zhang\cite{Zhang} discussed the effect of surface roughness on
the surface density of states of superfluid $^3$He-B.
Surface bound states in superfluid ${^3}$He-B are now recognized
as Majorana Fermions reflecting the topological property of the
bulk system.\cite{Schnyder,Qi,Okuda} A comprehensive
review from this aspect has been recently given by Mizushima et 
al.\cite{Mizushima} 

Experimental observation of the surface bound states in 
superfluid ${}^3$He was not performed
because of the lack of
appropriate surface probe for the neutral supefluid.
It has been recently reported,
\cite{Nagai2008,Aoki,Murakawa,Murakawa2011,Nagato2007} however, that the 
transverse acoustic impedance provides useful information on the
surface bound states in the B phase of superfluid ${}^3$He.  
The key point to identify the
surface bound states is   
the existence of the subgap in the surface density of 
states.\cite{Zhang,roughp,Vorontsov}
When the surface is specular, the surface density of states
is filled by the bound states up to the bulk energy gap $\Delta_{\rm b}$.
In the presence of rough surface, however, the bound state energies are
lowered and have a maximum $\Delta^*$ below the 
bulk energy gap $\Delta_{\rm b}$. 
As a result, there occurs a subgap between $\Delta^*$
and $\Delta_{\rm b}$. The pair excitations give rise to a typical cusp in the
real part of the transverse acoustic impedance and a peak in the imaginary
part at the frequency $\Delta^*+\Delta_{\rm b}$, which was observed by
Aoki et al.\cite{Aoki}

The subgap structure in the surface density of states of ${}^3$He-B 
was first reported by Zhang\cite{Zhang}, who
calculated the surface density of state using the quasi-classical
theory with thin dirty layer model. It was confirmed  by later
calculations using other rough surface models.\cite{roughp,Vorontsov}
However, the origin of the subgap structure has been a puzzle,
although Zhang\cite{Zhang} had suggested that it might be related to the
suppression of the parallel component of the order parameter near
the surface by the roughness. 
In Ref.\onlinecite{subgap}, we found the same subgap structure also 
in two-dimensional chiral superconducting state which has a full gap
at the Fermi surface as $^3$He-B. We showed that
the origin of the subgap is the level repulsion between the
bound state and the continuum states.

In this paper, we show that the subgap in the surface density of
states exists in superfluid $^3$He-B also in the presence of 
magnetic field perpendicular
to the surface and that the level repulsion is originated from
the structure of the bound state wave function 
that ensures the surface bound state to be Majonara Fermion.

This paper is organized as follows. 
In the next section, we discuss the possible form of the wave function
(Nambu amplitude)
of the surface bound state allowed by the symmetry of the Hamiltonian.
All the previous results\cite{Nagato2009,Chung,Volovik,Miz,Silaev} of the
bound state wave function agree with this possible form.
In section \ref{section:3}, we consider the quasi-classical Green
function for superfluid $^3$He-B with a plane surface and 
magnetic field perpendicular to the surface. Rough surface effects
are treated using the random $S$-matrix theory.\cite{rough,nagai_Verditz}
This theory is a self-consistent Born approximation theory with respect
to the surface scattering and includes a roughness parameter $W$ ($0\le
W\le 1$). The specular surface corresponds to $W=0$ and the fully
diffusive surface corresponds to $W=1$. We can show\cite{Nagai2008} that the random
$S$-matrix
theory with $W=1$ is equivalent to Ovchinnikov's rough surface boundary
condition.\cite{Ovchinnikov, Kopnin}
At the surface, we can separate the quasi-classical Green function into
two parts. One of them, which we shall call ``pole part'', includes
the contribution from the surface bound states and the other part,
which we shall call ``continuum part'', does not. From the structure
of the bound state wave function, we show that the ``pole part''
and the ``continuum part'' appear alternately in the multi-scattering
processes by the surface roughness. This leads to the level repulsion
between the bound state and the continuum states.   
In section~\ref{sec:4}, we consider a case with weak surface roughness.
Using perturbation theory with respect to the roughness parameter $W$,
 we show that the
bound state energies are lowered by the roughness and show how the upper edge
$\Delta^*$ appears.
Section~\ref{sec:5} is devoted to the self-consistent calculation of the
order parameters and the spin polarization density. To perform the
numerical calculations, we use the Riccati
representation of the quasi-classical Green function.
\cite{nagai_Verditz,Nagato1993,Higashitani,schopohl,eschrig} 
Using the self-consistent results, we calculate the surface density of
states and show
that the well defined subgap structure exists even in the presence of
magnetic
field.
In section~\ref{sec:6}, we  discuss the transverse acoustic impedance
and show how it depends on the magnetic field.
The final section is devoted to summary and discussion. In the
 Appendices, Riccati representation of $4\times 4$ quasi-classical 
Green function is discussed. Explicit expressions for the transverse
acoustic impedance is also presented.

\section{Surface Bound States}
Let us briefly review the surface bound states
of a superfluid ${}^3$He-B
filling $z>0$ domain with a plane
surface located at $z=0$.\cite{Nagai2008,Nagato2009, Chung, Volovik,Miz, Silaev} The magnetic field is applied in the
direction perpendicular to the plane surface.
We consider 4-dimensional Bogoliubov equation
\begin{equation}
 \int d\bm{r}' \mathcal{H}(\bm{r},\bm{r}')\Psi(\bm{r}')=E\Psi(\bm{r}),
\label{bdg}
\end{equation}
where $\Psi$ is the Nambu amplitude and 
the Hamiltonian of the system in 4-dimensional Nambu representation is
given by
\begin{equation}
\mathcal{H}(\bm{r},\bm{r}')=
 \begin{pmatrix}
  \xi(\nabla)\delta(\bm{r}-\bm{r}') & \Delta(\bm{r},\bm{r}')\\
  \Delta^\dagger(\bm{r}',\bm{r}) & -\xi(\nabla)\delta(\bm{r}-\bm{r}')
 \end{pmatrix}
 -h 
\begin{pmatrix}
 \sigma_z & \\ & -\sigma_z
\end{pmatrix} \delta (\bm{r}-\bm{r}'),
\end{equation}\label{energy}
where $\xi(\nabla)=\frac{\nabla^2}{2m^*}-\mu$ is the quasi-particle
energy
measured from the Fermi surface and $h$ is half the Larmor frequency 
$\omega_L$. 
\def\kp{\bm{K}}
\def\uk{u_{k}(z)}
The order parameter of $^3$He-B of the present system is
\begin{align}
\Delta(\bm{r},\bm{r}')&=\sum_{\bm{p}} e^{i\bm{p}\cdot(\bm{r}-\bm{r}')}
\Delta(z,\hat{\bm{p}}) \\
\Delta(z,\hat{\bm{p}})&=
\begin{pmatrix}
-d_x+i d_y & dz \\
dz & d_x+i d_y 
\end{pmatrix}
=
\begin{pmatrix}
- \Delta_t e^{-i\phi} & \Delta_\ell \\
\Delta_\ell &\Delta_t e^{i\phi},
\end{pmatrix}\\
\Delta_t&=\Delta_\parallel(z)\sin\theta, \\
\Delta_\ell&=\Delta_\perp(z)\cos\theta
\end{align}
where $\bm{d}$ is the $d$-vector, $\hat{\bm{p}}$ is a unit vector along
$\bm{p}$
and $\theta, \phi$  are the polar and the azimuthal angle of $\bm{p}$. 

Let us consider the bound state wave function. Since  the
momentum component $\bm{K}$ parallel to the surface is conserved when the surface
is specular, we consider the Nambu amplitude of the form
\begin{equation}
 \Psi(\bm{r})=e^{i\bm{K}\cdot\bm{r}}\sum_{\alpha=\pm}\Phi_\alpha(z)e^{i\alpha
  k z},
\end{equation}
with $k=\sqrt{p_F^2-\bm{K}^2}$ the $z$ component of the Fermi momentum.
Making use of the rotational symmetry around the $z$-axis, we rotate the
spin axis so that $\bm{d}$ is along the $y$ direction.
Applying the WKB approximation on Eq.(\ref{bdg}), we obtain the Andreev
equation
\begin{equation}
\mathcal{H}_\alpha \Phi_\alpha =
 \begin{pmatrix}
-i\alpha v_F\cos\theta\partial_z -h\sigma_z & \Delta_\alpha(z) \\
\Delta_\alpha^\dagger(z) & i\alpha v_F\cos\theta \partial_z+h \sigma_z  
 \end{pmatrix}
\Phi_\alpha=E\Phi_\alpha,\label{andreev}
\end{equation}
where $v_F=p_F/m^*$ is the Fermi velocity and
\begin{equation}
 \Delta_\alpha(z)=
\begin{pmatrix}
i \Delta_t & \alpha\Delta_\ell \\
\alpha\Delta_\ell & i \Delta_t
\end{pmatrix}\label{op}
\end{equation}
which satisfies
\begin{equation}
 \Delta_\alpha(z)^\dagger=-\Delta_{-\alpha}(z).
\end{equation}
 
The surface bound states can be obtained by seeking for damping
solutions
of Eq.(\ref{andreev}) and imposing the boundary condition $\Psi=0$ at $z=0$.
It is useful to note that $\mathcal{H}_\alpha$ satisfies
\begin{equation}
 \rho_2\sigma_1\mathcal{H}_\alpha\sigma_1\rho_2=\mathcal{H}_{-\alpha},
\end{equation}
where $\rho_i$'s and $\sigma_i$'s are Pauli matrices in the particle-hole
space
and the spin space, respectively. Since the bound state is non-degenerate,
it follows that
\begin{equation}
\Phi_-(z)=c\, \rho_2\sigma_1 \Phi_+(z), 
	    \end{equation}
where $c$ is a constant. Substituting this into the boundary condition
$\Phi_+(0)=-\Phi_-(0)$, we obtain
\begin{align}
\Phi_+(0)=
 u_1\begin{pmatrix} 1 \\ 0 \\ 0 \\ -ic \end{pmatrix}
 + u_2 \begin{pmatrix} 0 \\ ic \\ 1 \\ 0 \end{pmatrix}=-\Phi_-(0),\quad
c^2=1
\end{align}
where $u_1, u_2$ are constants. From the previous 
analyses,\cite{Nagato2009, Chung, Volovik,Miz, Silaev} one can conclude
that the damping solution is obtained when $c=1$ ( see below). At the surface,
therefore, the wave function
of the bound state is given by a linear combination of
\begin{equation}
 \Phi_\uparrow=\begin{pmatrix} 1 \\ 0 \\ 0 \\ -i \end{pmatrix},
\qquad \Phi\downarrow=\begin{pmatrix} 0 \\i \\ 1 \\ 0 \end{pmatrix}.
\label{majorana}
\end{equation}

When we assume  that 
$\Delta_\parallel$ and $\Delta_\perp$ are constant, we can find
explicit solutions for the bound 
state.\cite{Nagato2009, Chung}
In the absence of magnetic field.
we can find both positive and negative energy bound state
for each Fermi momentum $\bm{p}_{\mathrm{F}}=(\bm{K}, k)$.
For positive energy eigen value $E=\Delta_\parallel \sin\theta$,
 the
eigen function is given by  
\begin{equation}
 \Psi_{\bm{K}}^{(+)}(\bm{r})=e^{i\bm{K}\cdot\bm{r}}\sin(kz)e^{-\kappa z}\left(\Phi_\uparrow -
e^{i{\phi}}\Phi_\downarrow\right)\label{psiplus}
\end{equation}
and for negative energy eigen value $E=-\Delta_\parallel \sin\theta$
\begin{equation}
 \Psi_{\bm{K}}^{(-)}(\bm{r})=e^{i\bm{K}\cdot\bm{r}}\sin(kz)e^{-\kappa z}
\left(e^{-i{\phi}}\Phi_\uparrow  +\Phi_\downarrow\right).\label{psiminus}
\end{equation}
with $\kappa=\Delta_\perp/v_F$. Here we have rotated the spin axis back
to the original one.
\noindent
In the presence of magnetic field perpendicular to 
the surface,
the bound state energy  has a gap 
\begin{equation}
 E=\pm\sqrt{\Delta_t^2+\Delta_Z^2},\ \ \Delta_Z=h.\label{b_energy}
\end{equation}
We call $\Delta_Z$ ``Zeeman gap'' in this paper.
The corresponding eigen function is given by a linear combination
\begin{equation}
  a \Psi_{\bm{K}}^{(+)}(\bm{r})+ b \Psi_{\bm{K}}^{(-)}(\bm{r})
\end{equation}
and $a,b$ are determined by
\begin{equation}
 \frac{a}{-he^{-i\phi}}=\frac{b}{E-\Delta_t}
\end{equation}

The Nambu amplitudes $\Phi_\uparrow$ and
$\Phi_\downarrow$ are the eigen functions of the zero energy surface
bound states of the polar state ($\Delta_t=0$). 
Treating $\Delta_t$ and $h$ as perturbation,\cite{Volovik, Silaev} 
one can also show that the bound state in the BW state is given by
the linear combination of $\Phi_\uparrow$ and $\Phi_\downarrow$. 
It is to be noted that, in the Nambu amplitudes $\Phi_\uparrow$ and
$\Phi_\downarrow$, the particle component and the hole component enter
with the same weight, which is a basis for the surface bound states
to be  Majorana Fermions. 

\section{Quasi-classical Green Function in the presence of Magnetic Field}\label{section:3}
In this section, we discuss the quasi-classical Green function of
$^3$He-B with the plane surface in the $z=0$ plane and the magnetic
field 
parallel to the $z$ axis.

We consider the quasi-classical Green function $\hat{G}_\alpha(\epsilon,z)\ (\alpha=\pm1)$ for the
Fermi momentum $\bm{p}_\alpha=(\bm{K}, \alpha k)$. $\bm{p}_+$ and
$\bm{p}_-$ are
the
outgoing and  the incoming Fermi momentum, respectively.
Since the
system
keeps rotational symmetry around the $z$ axis, we consider $\hat{G}_\alpha$
for
$\phi=\pi/2\ ((\bm{K})_x=0)$. 
Green's function for arbitrary $\phi$ is obtained by
\begin{equation}
 \mathcal{R}^\dagger(\phi)\hat{G}_\alpha \mathcal{R}(\phi), \quad
 \mathcal{R}(\phi)=\exp\left(i\left(\phi-\pi/2\right)\rho_3\sigma_3/2\right)
\label{rphi}
\end{equation}
The Eilenberger equation for $\hat{G}_\alpha$ is now given by
\begin{align}
{v_F\cos\theta}\partial_z \hat{G}_\alpha&={i\alpha}
\left[\hat{G}_\alpha, \mathcal{E}_\alpha(z)\right] \label{Eilenberger}\\
\mathcal{E}_{\alpha}&=\mathcal{E}_0+\mathcal{E}_h \label{EE}\\
\mathcal{E}_0&=\epsilon\rho_3
+\begin{pmatrix}
   & \Delta_\alpha \\
\Delta_{-\alpha} & 
 \end{pmatrix}\\
\mathcal{E}_h&=h\sigma_3\rho_0,
\end{align}
where the order parameter $\Delta_\alpha$ is given by Eq. (\ref{op}).

In the bulk region far from the surface, $\hat{G}_\alpha$ is given by\cite{BZ,Ashida_Nagai}
\begin{align}
\hat{G}_\alpha^{(0)}=\sum_{\sigma=\pm 1}i P_\sigma
\frac{\mathcal{E}_\alpha}{E_\sigma},\label{bulk} 
\end{align}
where $P_\sigma$ is a projection operator
\begin{align}
 P_\sigma&=\frac{1}{2}\left(1-\sigma\frac{Q}{|Q|}\right),\\
  Q&=\mathcal{E}_0\mathcal{E}_h+\mathcal{E}_h\mathcal{E}_0,\\
  |Q|&=\sqrt{{Q}^2}=2h\sqrt{\epsilon^2-\Delta_t^2},
\end{align}
and 
\begin{align}
 E_\sigma=\sqrt{\epsilon^2-\Delta_t^2-
\Delta_\ell^2+h^2-\sigma Q}\label{q_particle}
\end{align}
In the above, we have defined square root such that has positive
imaginary
part.

When the surface is specular, $\hat{G}_+$ and $\hat{G}_-$ should satisfy the
boundary condition at the surface $z=0$
\begin{align}
 \hat{G}_+(0)=\hat{G}_-(0)=\hat{G}_S.\label{boundary}
\end{align} 
Using Eqs.(\ref{Eilenberger}), (\ref{bulk}) and (\ref{boundary}), we can
determine the quasi-classical Green function of the system with 
specular surface
(see Appendix \ref{app2}).  
The quasi-classical Green function of present definition
satisfies the normalization condition $\hat{G}_\alpha^2=-1$.

From the
symmetrical property of $\mathcal{E}_\alpha$
\begin{equation}
 \rho_2\sigma_1 \mathcal{E}_\alpha \sigma_1 \rho_2=-\mathcal{E}_{-\alpha},
\end{equation} 
we can show that
\begin{equation}
 \rho_2\sigma_1 \hat{G}_\alpha (z) \sigma_1
  \rho_2=-\hat{G}_{-\alpha}(z),
\label{g:symmetry}
\end{equation}
therefore
\begin{equation}
 \rho_2\sigma_1 \hat{G}_s \sigma_1 \rho_2=-\hat{G}_s.\label{symmetry}
\end{equation}

Let us consider how the surface bound states contribute to the 
quasi-classical Green function. For that purpose,
it is useful to remind that $\hat{G}_s$ is
expressed
in terms of the eigen functions $\Phi_{n}$ of the Andreev equation 
(\ref{andreev})
in a form\cite{rough,nagai_Verditz,aahn}
\begin{align}
 \hat{G}_s=\dfrac{-1}{2v_F\cos\theta}\frac{1}{2}\rho_3\left(
\sum_n \left. \dfrac{\Phi_{n}(z)\Phi_{n}^\dagger(0)}
{\epsilon-E_n}\right|_{z=+0}
+\sum_n \left. \dfrac{\Phi_{n}(z)\Phi_{n}^\dagger(0)}
{\epsilon-E_n}\right|_{z=-0}
\right).
\end{align}
Bearing this in mind, we prepare 4 mutually orthogonal Nambu amplitudes
\begin{align}
 |\Phi_1>=\frac{1}{\sqrt{2}}\Phi_\uparrow,\quad
 |\Phi_2>=\frac{1}{\sqrt{2}}\Phi_\downarrow,\quad
 |\Phi_1'>=\rho_3\Phi_1,\quad
 |\Phi_2'>=\rho_3\Phi_2.\label{base}
\end{align}
When we choose them as base vectors and calculate the matrix elements of 
$g_s=\rho_3 \hat{G}_s$, 
we find that $g_s$ is block diagonalized, i.e.,
\begin{align}
 g_s=& \rho_3 \hat{G}_s\nonumber\\
    =&
\sum_{i,j=1,2}
\left(|\Phi_i><\Phi_i|g_s|\Phi_j><\Phi_j|
+|\Phi_i'><\Phi_i'|g_s|\Phi_j'><\Phi_j'|\right)\\
 \equiv& \sum_{i,j=1,2}
\left(|\Phi_i>g^{(1)}_{ij}<\Phi_j|
+|\Phi_i'>g^{(2)}_{ij}<\Phi_j'|\right).
\end{align}
This can be proved by noting that  $\rho_2\sigma_1|\Phi_1>=-|\Phi_1>$,
$\rho_2\sigma_1|\Phi_2>=-|\Phi_2>$ together with Eq.(\ref{symmetry}).  
\begin{align}
  <\Phi_i|g_s|\Phi_j'>&=<\Phi_i|\rho_3 \hat{G}_s|\Phi_j'>\nonumber\\
&=-<\Phi_i|\rho_3\rho_2\sigma_1 \hat{G}_s\sigma_1\rho_2\rho_3|\Phi_j>\nonumber\\
&=-<\Phi_i|\rho_2\sigma_1\rho_3 \hat{G}_s \rho_3\sigma_1\rho_2|\Phi_j>\nonumber\\
&=-<\Phi_i|\rho_3 \hat{G}_s \rho_3|\Phi_j>=-<\Phi_i|\rho_3 \hat{G}_s|\Phi_j'>=0.
\end{align}
As was emphasized in the previous section, the bound state wave
functions at the surface $z=0$ are given by the linear combination of
$|\Phi_1>$ and $|\Phi_2>$. It follows that the bound states appear
in $g^{(1)}$, while $g^{(2)}$ has no
contribution from the bound states at all.
We call $g^{(1)}$ the ``pole part'' and $g^{(2)}$ the
``continuum part'', although $g^{(1)}$ has contribution
also from the continuum states.  
From the normalization condition $\hat{G}_s^2=-1$ we find
\begin{equation}
 g^{(1)} g^{(2)}=-1.\label{normalization2}
\end{equation}

Let us now consider the rough surface effects using the random S-matrix
model.\cite{rough,roughp,nagai_Verditz}. According to the theory, the
Green functions at the surface are given by
\begin{align}
 \hat{G}_\pm&=\hat{G}_s+(\hat{G}_s \pm i)\mathcal{\hat{G}}(\hat{G}_s \mp i),\\
    \mathcal{\hat{G}}&=\frac{1}{\hat{G}_s^{-1}-\Sigma},
\end{align}
where $\Sigma$ is the surface self energy given by
\begin{align}
 \Sigma&=2W\left< \frac{1}{2}\left(\mathcal{\hat{G}}+
    \rho_3\sigma_3\mathcal{\hat{G}}\sigma_3\rho_3\right)\right>,\label{self}\\
    <\cdots>&=2\int_0^{\pi/2}d\theta\sin\theta\cos\theta \cdots.
\end{align}
Here, $W (0\le W \le 1)$ is a parameter which specifies the roughness of 
the surface;
$W=0$ corresponds to the specular surface and $W=1$ to the completely
diffusive surface.\cite{rough,roughp,nagai_Verditz} 
When $W=1$, the random S-matrix model is equivalent to Ovchinnikov's
boundary condition.\cite{Nagai2008,Ovchinnikov, Kopnin}
Note that Eq.(\ref{self}) is
slightly
different from our previous definition 
$\Sigma=2W<\mathcal{\hat{G}}>$\cite{rough,roughp}. This is
because
the integral over the azimuthal angle $\phi$ has been already performed
(see Eq. (\ref{rphi})).

From the definition of
Eq.~(\ref{self}) (see also Eq.~(\ref{sigma:form})), we can show that $\Sigma$ is
also block diagonalized by the same base vectors as
\begin{align}
 \Sigma=\rho_3 \sum_{i,j=1,2}
\left(|\Phi_i>s^{(1)}_{ij}<\Phi_j|
+|\Phi_i'>s^{(2)}_{ij}<\Phi_j'|\right),
\end{align}
where $s^{(1)}$ and $s^{(2)}$ are $2\times 2$ diagonal matrices.
We can also show that
\begin{align}
 \mathcal{G}=\dfrac{1}{\hat{G}_s^{-1}
 -\Sigma}
&=\rho_3\sum_{ij}\left(
|\Phi_i>\left(\frac{1}{(g^{(1)})^{-1}-s^{(2)}}\right)_{ij}<\Phi_j|+|\Phi_i'>
\left(\frac{1}{(g^{(2)})^{-1}-s^{(1)}}\right)_{ij}<\Phi_j'|\right),\label{gsep}
\end{align}
where we used Eq. (\ref{normalization2}) and $|\Phi_i'>=\rho_3 |\Phi_i>$.
The self energy equation (\ref{self}) is also separated.
\begin{align}
s^{(1)}=&2W\left< \left(\frac{1}{(g^{(1)})^{-1}-s^{(2)}} \right)_
\mathrm{diagonal}\right>\label{self(1)}\\
s^{(2)}=&2W\left< \left(\frac{1}{(g^{(2)})^{-1}-s^{(1)}} \right)_
\mathrm{diagonal}\right>
\end{align}
The first term in Eq. (\ref{gsep}) shows how the bound state energy
is modified by the surface roughness. In fact, when we formally expand 
the first term, we find that
\begin{equation}
 g^{(1)}+g^{(1)}s^{(2)}g^{(1)}+g^{(1)}s^{(2)}g^{(1)}s^{(2)}g^{(1)}+\cdots,
\end{equation}
namely, the ``pole part''
and the ``continuum part'' appear alternately in the multi-scattering 
processes by the roughness. It implies that there occurs level 
repulsion between the
bound state and the continuum states. It leads to the
existence of a maximum  $\Delta^*$ of the bound state energy
and a subgap between $\Delta^*$ and the bulk energy gap. 
We have shown in Ref.\onlinecite{subgap} that this actually happens
in the two-dimensional chiral superconductor.

Finally we consider the surface density of States $D(\epsilon)$ 
which is calculated from\cite{roughp}
\begin{align}
 D(\epsilon)&=\int_0^{\pi/2}d\theta\, \sin\theta \,n(\epsilon,\theta),
\label{total_density}\\
    n(\epsilon,\theta)&= \frac{1}{4}N(0)\mathrm{Im}\, \mathrm{tr}\left[
\rho_3\frac{1}{2}(\hat{G}_+(\epsilon+i0)+\hat{G}_-(\epsilon+i0))\right]
\label{angle_resolved}\\
 &= \frac{1}{4}N(0)\mathrm{Im}\, \mathrm{tr}\left[
\frac{1}{(g^{(1)})^{-1}-s^{(2)}}
             +\frac{1}{(s^{(1)})^{-1}-g^{(2)}}
+\frac{1}{(g^{(2)})^{-1}-s^{(1)}}
             +\frac{1}{(s^{(2)})^{-1}-g^{(1)}}\right],
\end{align}
where $N(0)$ is the normal state density of states.
\section{Perturbation Theory}\label{sec:4}

The advantage of the random $S$-matrix model is that we can treat
the surface roughness in a unified way from  the specular limit
to the diffusive limit.
In this section, we consider the surface density of states when the
surface
roughness is weak, i. e., $W$ is small. 

For simplicity, we assume in
this
section that the order parameters $\Delta_\parallel, \Delta_\perp$
are spatially constant. In that case, $\hat{G}_s$ at the specular surface is 
given by\cite{rough,roughp,nagai_Verditz}
\begin{align}
 \hat{G}_s&=\hat{G}_+^{(0))}+(\hat{G}_+^{(0))}+i)\frac{-1}{\hat{G}_+^{(0))}+\hat{G}_-^{(0))}}(\hat{G}_+^{(0))}-i)\\
&=\hat{G}_-^{(0))}+(\hat{G}_-^{(0))}-i)\frac{-1}{\hat{G}_+^{(0))}+\hat{G}_-^{(0))}}(\hat{G}_-^{(0))}+i).\label{specularG}
\end{align}
Substituting Eq.(\ref{bulk}), we obtain expressions for
 $\hat{G}_s$ as is shown in Appendix \ref{app1}. We can also calculate
 explicitly
$g^{(1)}$ and $g^{(2)}$ and find that $g^{(1)}$ has poles
at
\begin{equation}
 \epsilon=\pm \sqrt{\Delta_t^2+h^2},
\end{equation}
while $g^{(2)}$ has no pole.  In fact,  we
obtain a simple result 
for the traces
 \begin{align}
  \mathrm{tr}\, g^{(1)}&=\epsilon \dfrac{i E_+\left(1+\frac{h}{\sqrt{\epsilon^2-\Delta_t^2}}\right)+i E_-\left(1-\frac{h}{\sqrt{\epsilon^2-\Delta_t^2}}\right)-2\Delta_\ell}{\epsilon^2-(\Delta_t^2+h^2)}\label{tr(1)}\\
  \mathrm{tr}\, g^{(2)}&=\epsilon \dfrac{i E_+\left(1+\frac{h}{\sqrt{\epsilon^2-\Delta_t^2}}\right)+i E_-\left(1-\frac{h}{\sqrt{\epsilon^2-\Delta_t^2}}\right)+2\Delta_\ell}{\epsilon^2-(\Delta_t^2+h^2)}.
 \end{align}

Let us consider the angle resolved surface density of
states $n(\epsilon,\theta)$ in low energy region.
When $W$ is small, the pole part contribution to 
$n(\epsilon,\theta)$ may be written as
\begin{equation}
 n(\epsilon,\theta)=\frac{N(0)}{4}\mathrm{Im}\,\mathrm{tr}
\left(
\frac{1}{(g^{(1)})^{-1}-s^{(2)}}+s^{(1)}
\right) \label{dens2}
\end{equation}
When the surface is specular, $W=0$, therefore $s^{(1)}=s^{(2)}=0$. 
We find from Eq.(\ref{tr(1)}) the bound state peaks
\begin{align}
 n(\epsilon,\theta)=&\pi N(0) \Delta_\ell |\epsilon|
  \delta\left(\epsilon^2-(\Delta_t^2+h^2)
\right)\nonumber\\
=&\frac{\pi N(0) \Delta_\ell}{2}
\left(\delta\left(\epsilon-\sqrt{\Delta_t^2+h^2}\right)+
\delta\left(\epsilon+\sqrt{\Delta_t^2+h^2}\right)
\right)\label{W_zero_peak}
\end{align}
and also a continuous spectrum of the Bogoliubov quasi-particles
which begins from the minimum energy $\Delta_\perp-h$.\cite{Ashida_Nagai}
The first correction by $W$ comes from $s^{(1)}$.
\begin{equation} 
 \mathrm{Im}\,\mathrm{tr}\,s^{(1)}=\mathrm{Im}\,\mathrm{tr}\, 2W <g^{(1)}>
=8\pi W |\epsilon|\frac{\Delta_\perp}{\Delta_\parallel^2}
\sqrt{1-\frac{\epsilon^2-h^2}{\Delta_\parallel^2}}\Theta(|\epsilon|-h),
\label{s1correction}
\end{equation}
where $\Theta$ is the theta function. This describes the effects by
the scattering of the bound states by the surface roughness.

\begin{figure}[h]
 \begin{center}
\includegraphics{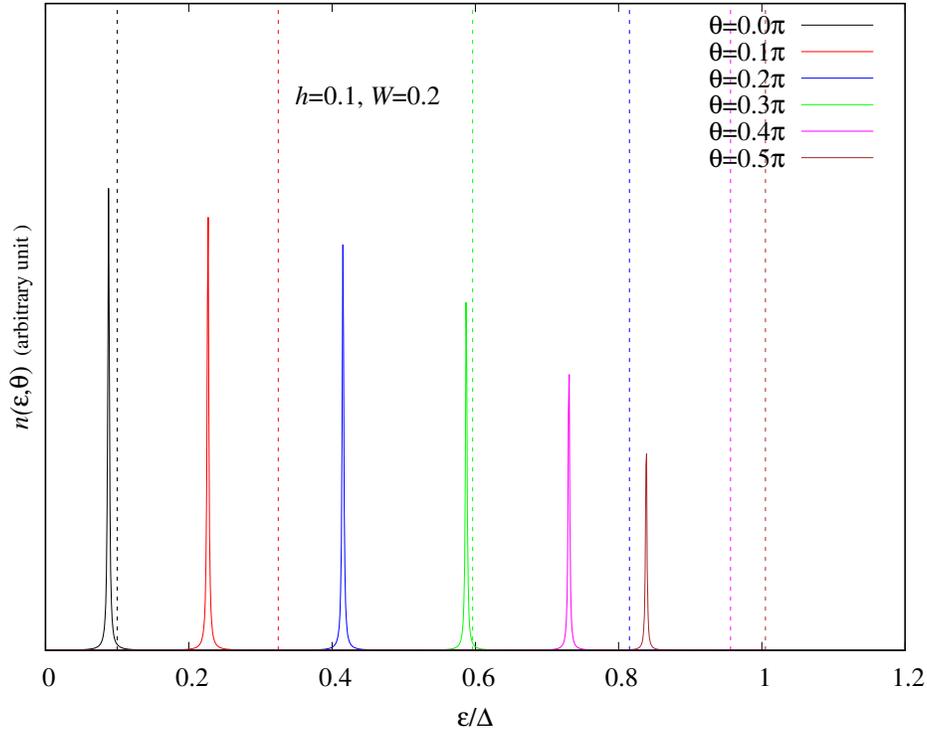}  
 \end{center}
\caption{\label{fig:fig1}Bound states contribution to the density of state when
 $h=0.1\Delta$
 and $W=0.2$. Artificial width is given to the peaks to show the
$\theta$ dependence of the spectral weight. The vertical broken lines
show the bound state energies when $W=0$.}
\end{figure}

Now we consider the correction by $W$ to the the first term of
Eq. (\ref{dens2}), putting $s^{(2)}=2W< g^{(2)}>$, to study how the 
bound state energy is modified by the
surface roughness. 
\begin{equation}
 n(\epsilon,\theta)=\frac{N(0)}{4}\mathrm{Im}\,\mathrm{tr}
\left(
\frac{1}{(g^{(1)})^{-1}-2W<g^{(2)}>}
\right) 
\end{equation}
Since the numerical
calculation is necessary to evaluate $<g^{(2)}>$
when the magnetic field is finite, we further
simplify the problem by assuming $\Delta_\parallel=\Delta_\perp=\Delta$.
The result is shown in Fig. \ref{fig:fig1}.

Figure \ref{fig:fig1} shows the bound state contribution to $n(\epsilon,\theta)$
when $h=0.1\Delta$ and $W=0.2$. Bound state peaks are artificially
broadened to show the $\theta$ dependence of the spectral weight.
The vertical broken line of the same color shows the bound state
energy Eq. (\ref{b_energy}) when $W=0$. We find that the bound state
energies are shifted to lower energy by the level repulsion with the
continuum states. 
The Zeeman gap $\Delta_Z$ is  lowered by
the roughness. The bound state at $\theta=\pi/2$, which has the
maximum energy, has also a lower energy than that in the case of $W=0$. 
This energy is $\Delta^*$ at the present level of approximation. 
Important point is that the spectral weight of
the bound state peak at $\Delta^*$ is finite in contrast
to the case of specular surface where the spectral weight is
zero when $\theta=\pi/2$ (see Eq. (\ref{W_zero_peak})).

The presence of $\Delta^*$ should be taken into account
in the calculation of $s^{(1)}$ of Eq. (\ref{self(1)}).
Equation (\ref{s1correction}) of $s^{(1)}$ was obtained
by integrating the pole contribution from the bound states over $\theta$.
When $W=0$ the bound state energies extend up to $\Delta_\parallel$. But 
now they are limited below $\Delta^*$.
Moreover the spectral weights of the bound states are
finite below $\Delta^*$. It follows that there occurs a finite
jump to zero in Im tr $s^{(1)}$ at $\Delta^*$. Bound state
contributions to the density of states are, thus, confined in the
energy range $\Delta_Z < \epsilon < \Delta^*$, which gives rise to
subgap structure,

In addition to the bound states, there is contribution to
the density of states from the Bogoliubov quasi-particle
excitations which have a minimum energy $\Delta_\perp-h$.\cite{Ashida_Nagai}   
When $W$ is sufficiently small, $\Delta^*$ may merge into this
excitation continuum.

In this section, we have discussed using a simplified model.
In the following sections, we show the results of more realistic
calculations based on the self-consistent order parameter.

\section{Self-Consistent Order Parameter, Spin Polarized Density and Surface Density of States}\label{sec:5}

In this section we discuss the order parameters, the spin polarization
density and the surface density of states of ${}^3$He-B  under the
magnetic field normal to the rough surface. 

The self-consistent order parameter $\Delta(z)$ and the spin
polarization density $S_z(z)$ are determined
by solving the gap equation 
\begin{align}
 \Delta(z,{\hat p}) &= \frac{
 \displaystyle
 3\pi T \sum_{\omega_n} \int_0^{2\pi}\frac{d\phi_k}{2\pi}
 \sum_\alpha^{\omega_c}\int_0^{\pi/2}\frac{d\theta_k}{2}\sin\theta_k
 ~{\hat p}\cdot{\hat k}_{\alpha}
 \left.{\hat G}_{\alpha}(K,z,i\omega_n)\right|_{12}
 }{
 \displaystyle
 \ln\frac{T}{T_c}+\sum_{\omega_n>0}^{\omega_C}\frac1{n+1/2}
 } \label{eq:gapeq}
\end{align}
and the spin polarization equation
\begin{align}
 S_z(z) &= N(0)\tilde{h}  - \frac{\pi TN(0)}4
 \sum_{\omega_n}^{\omega_c}\int_0^{\pi/2}
 d\theta\sin\theta \sum_\alpha{\rm tr}_{\rm spin}
 \left[\sigma_3\left.\hat
 G_{\alpha}(K,z,i\omega_n)\right|_{11}\right]
 \label{eq:spineq}
\end{align}
where the subscript $11$ and $12$ mean the $11$ and $12$ element of
the quasi-classical Matsubara Green's function $\hat G(i\omega_n)$
in particle-hole space, respectively.
The Fermi liquid correction by $F_0^a$ to
the effective magnetic field is given by 
\begin{align}
 \tilde{h}=\frac{\tilde\omega_L}{2} &=h - \frac{F_0^a}{N(0)}S_z\nonumber\\
 &=h \left( 1 - \frac{F_0^a}{1+F_0^a} \left(\frac{S_z}{S_N}\right)\right)
 \label{eq:tildewl}
\end{align}
where $h=\omega_L/2=\gamma H/2$ is half the external Larmor frequency. 

In order to solve the above equations numerically, we employ 
the Riccati representation of the quasi-classical Green function.
\cite{rough,roughp,schopohl,eschrig} Details are discussed
 in Appendix~\ref{app2}. 
We obtain the self-consistent order parameter and the spin polarization
density by solving the above equations iteratively. For given
$\Delta_\parallel$, $\Delta_\perp$ and $S_z$, we solve the Riccati
equations and the surface self energy equations. Using the obtained
self energy $\Sigma$, we calculate the gap equation and the spin
polarization density, and use the results of $\Delta_\parallel, \Delta_\perp$ and
$S_z$ for the next step.

\begin{figure}[h]
\begin{center}
 \includegraphics[width=0.49\columnwidth]{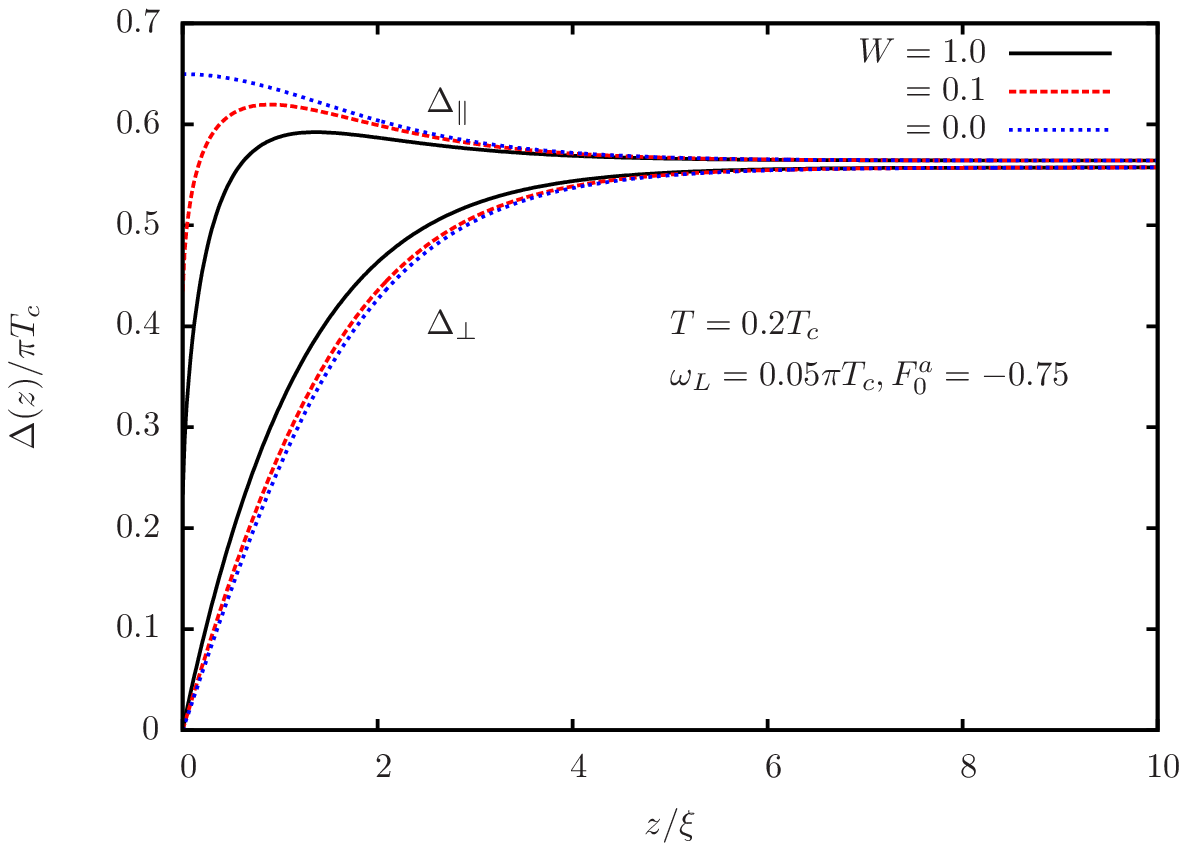}
 \includegraphics[width=0.49\columnwidth]{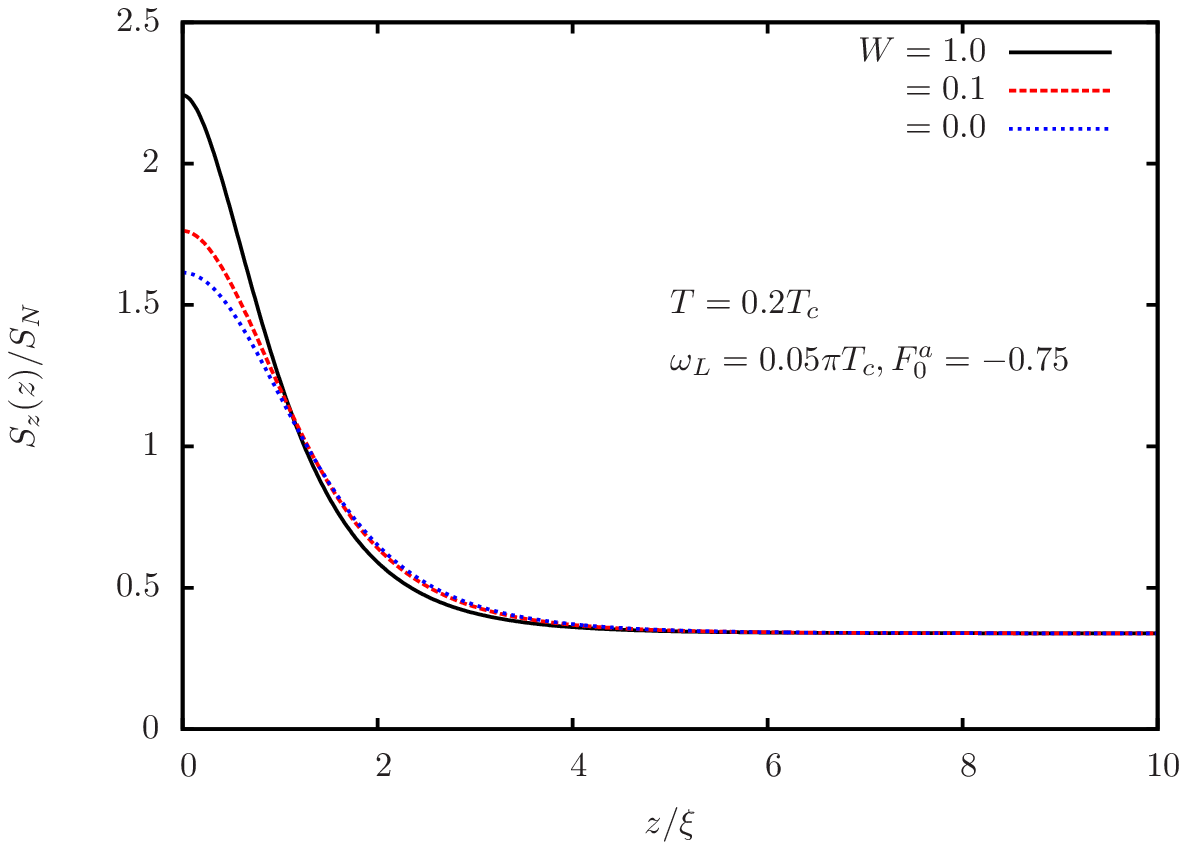}
 \end{center}
\caption{\label{fig:deltaspin4W} Self-consistent  order parameter 
 $\Delta(z)$ and  spin polarization density $S_z(z)$ 
for some typical values of the
 roughness parameter $W$ are plotted 
against the distance from the surface scaled by 
 the coherence length $\xi=v_F/\pi T_c$ . $T=0.2T_c$ and $\omega_L=0.05\pi T_c$}
\end{figure}

Typical results are shown in Fig. \ref{fig:deltaspin4W}.
We have chosen the temperature $T=0.2T_c$, the cut-off frequency 
$\omega_c=30T_c$, the external Larmor frequency 
$\omega_L=0.05\pi T_c$, which is about $0.185$T when $T_c=1.828$mK,
and the Fermi liquid parameter $F_0^a=-0.75$. 
The roughness parameter $W=1$ corresponds to the diffusive surface 
and $W=0.0$ corresponds to the specular surface. 

In the bulk region the perpendicular component $\Delta_\perp$ is 
smaller than the parallel component $\Delta_\parallel$,
since the magnetic field is applied perpendicular
to the surface. 
In the vicinity of the surface the profiles of the order parameter
are similar to that of the BW state  without  magnetic field 
discussed in our previous paper.\cite{roughp} 
The enhancement of the spin polarization density is seen at the
surface. At lower temperatures, the enhancement is most pronounced
in the diffusive limit $W=1$. At higher temperatures, however,
dependence of the profile on $W$ becomes small.

\begin{figure}[h]
\begin{center}
 \includegraphics{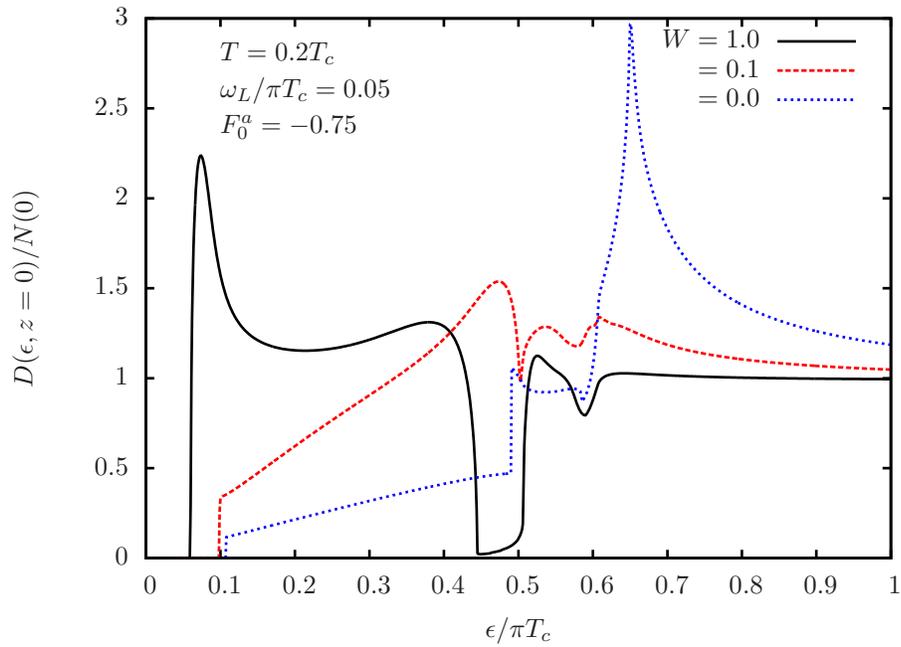}
\end{center}
\caption{\label{fig:densW} $W$ dependence of the surface density of states  
at $T=0.2T_c$ under the magnetic field $\omega_L= 0.05\pi T_c$.}
\end{figure}

The surface density of states is 
calculated from Eqs.~(\ref{total_density}) and (\ref{angle_resolved}).
In Fig.~\ref{fig:densW} we show the $W$ dependence of the 
surface density of states of
${}^3$He-B at $T=0.2T_c$
calculated using the self-consistent order parameter and the spin
density given in Fig. \ref{fig:deltaspin4W}. $W=0$ corresponds to the
specular surface, $W=1$ to the diffusive limit and $W=0.1$ corresponds
to
the surface with specularity factor about 0.5.\cite{roughp} 
We find that the ``Zeeman gap''  $\Delta_Z$ decreases
with $W$, 
as discussed in the previous section. 
The upper edge $\Delta^*$ of the surface bound states band 
can be seen for any $W\ne 0$. When $W=1$, well defined subgap structure
exists. At $W=0.1$, however, 
$\Delta^*$ is merged into the
continuum spectrum of the Bogoliubov quasi-particle excitations and
subgap structure is no longer found.  

\begin{figure}
\begin{center}
 \includegraphics{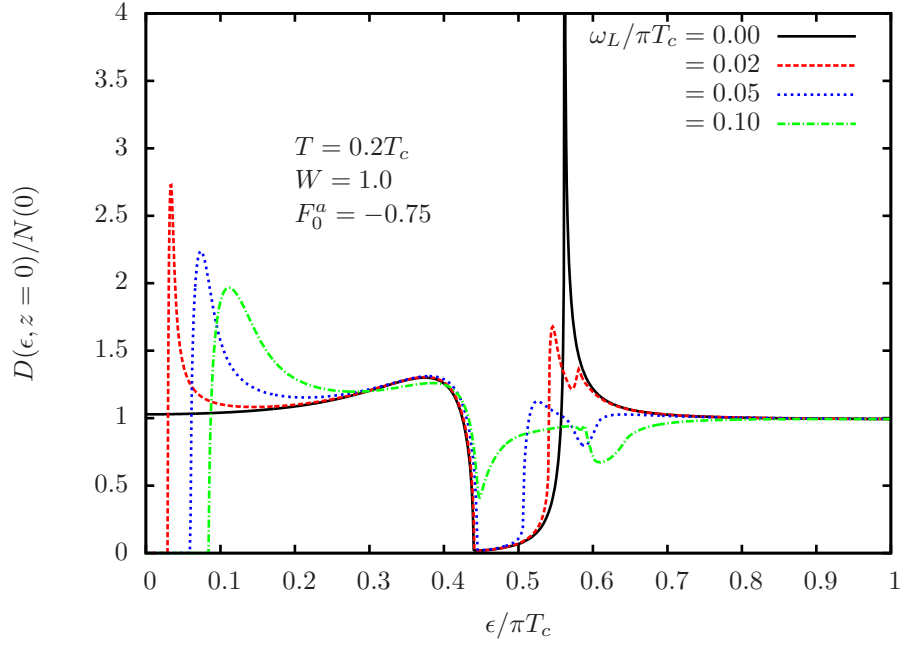}
\end{center}
\caption{\label{fig:denswl} Surface density of states at $T=0.2T_c,
 W=1$ under the magnetic fields 
 $\omega_L=0.0$, $0.02$, $0.05$, and $0.10 \pi T_c$ . }
\end{figure}

Magnetic field dependence of the surface density of states is shown in
Fig.~ \ref{fig:denswl}. We show the surface density of states
at $T=0.2T_c$ and $W=1$ under the external magnetic field 
$\omega_L/\pi T_c=0.0$, $0.02$, $0.05$ and $0.10$.
The ``Zeeman gap'' $\Delta_Z$ increases with
the magnetic field, as it should be. 
On the other hand the upper edge $\Delta^*$ 
does not so much depend on the magnetic field. 
The sharp peak at the the bulk energy gap ($\sim 0.5\pi T_c$)
that exists when $\omega_L=0$ 
is splitted by the magnetic field. This reflects the Zeeman splitting
of the Bogoliubov excitations.

\begin{figure}[h]
\begin{center}
 \includegraphics{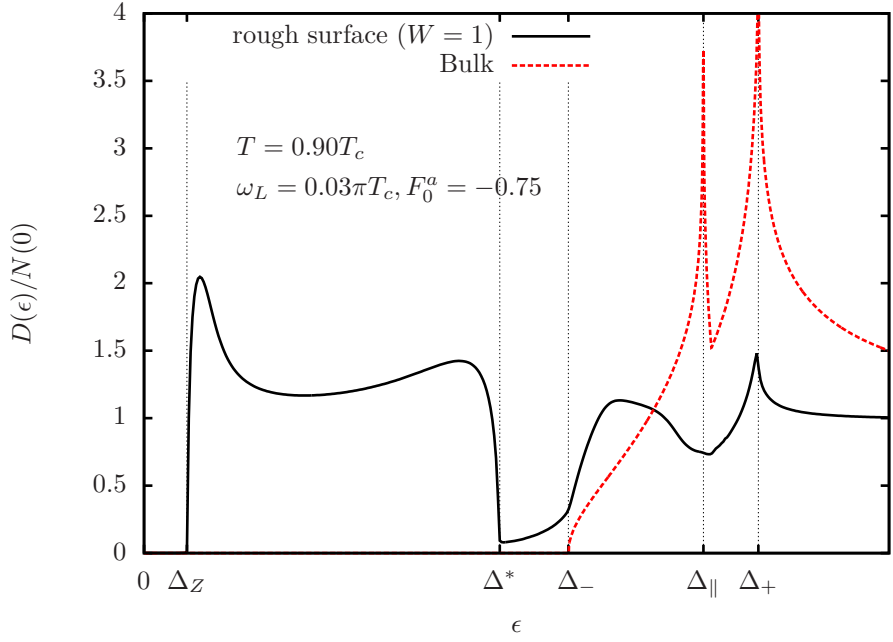}
\end{center}
\caption{\label{fig:denschar} Characteristic energies $\Delta_Z,
 \Delta^*, \Delta_-$, and $\Delta_+$ found in the surface density of states.
 The solid curve shows the local density of states at the rough surface
 when $T=0.9T_c, \omega_L=0.03\pi T_c$. 
 The dashed curve is the density of states in the bulk region. 
 }
\end{figure}

As can be seen in Figs. \ref{fig:densW} and \ref{fig:denswl},
the surface density of states is characterized by some band edges 
as well as some peaks. In Fig.\ref{fig:denschar}, we plot the
surface density of states when $W=1$, $T=0.9T_c$ and $\omega_L=0.03\pi
T_c$
together with the bulk density of states. We show the locations of such
singularity energies  $\Delta_Z$, $\Delta^*$, $\Delta_-$ and $\Delta_+$. 
Let us discuss on the behavior of those characteristic energies.
The energy $\Delta^*$, the upper edge of the surface
bound state band, 
increases toward the bulk energy gap as $W$ decreases.\cite{roughp} 
On the other hand, $\Delta^*$ depends little 
on the magnetic field normal to the surface.

The energies $\Delta_+$ and $\Delta_-$ are related to the Zeeman 
splitting of the bulk energy gap. 
From the quasi-particle energy  
of the bulk BW state under the magnetic field,\cite{Ashida_Nagai} 
 we can find
\begin{align}
 \Delta_+&=\begin{cases}\displaystyle
	   \Delta_{\perp b}+\frac{\tilde\omega_{Lb}}2 & \text{for}~ \Delta_{\parallel b}^2-\Delta_{\perp b}^2 <
	    \displaystyle\frac{\tilde\omega_{Lb}}2\Delta_{\perp b}\\
	   \Delta_{\parallel b}\sqrt{\displaystyle
	    1+\frac{\tilde\omega_{Lb}^2}4\frac1{\Delta_{\parallel b}^2
	    -\Delta_{\perp b}^2}}
	    & \text{for}~ \Delta_{\parallel b}^2-\Delta_{\perp b}^2 >
	    \displaystyle\frac{\tilde\omega_{Lb}}2\Delta_{\perp b} 
	   \end{cases}\\
 \Delta_-&=\Delta_{\perp b}-\frac{\tilde\omega_{Lb}}2 & 
\end{align}
where the subscript $b$ means the bulk value.
As the magnetic field increases, $\Delta_-$ decreases. 
Therefore, when the magnetic field $\omega_L$ becomes sufficiently large, 
$\Delta_-$ becomes less than $\Delta^*$ and the subgap between $\Delta^*$
and $\Delta_-$ disappears. This also happens when $W$ is small.
When the magnetic field is weak and $W$ is large, 
the surface states are characterized by the four energies
$\Delta_Z$, $\Delta^*$, $\Delta_-$ and $\Delta_+$.
In other cases,
the surface states are mainly characterized 
by the two energies $\Delta_Z$ and $\Delta_+$.

\begin{figure}
\begin{center}
 \includegraphics[width=0.49\columnwidth]{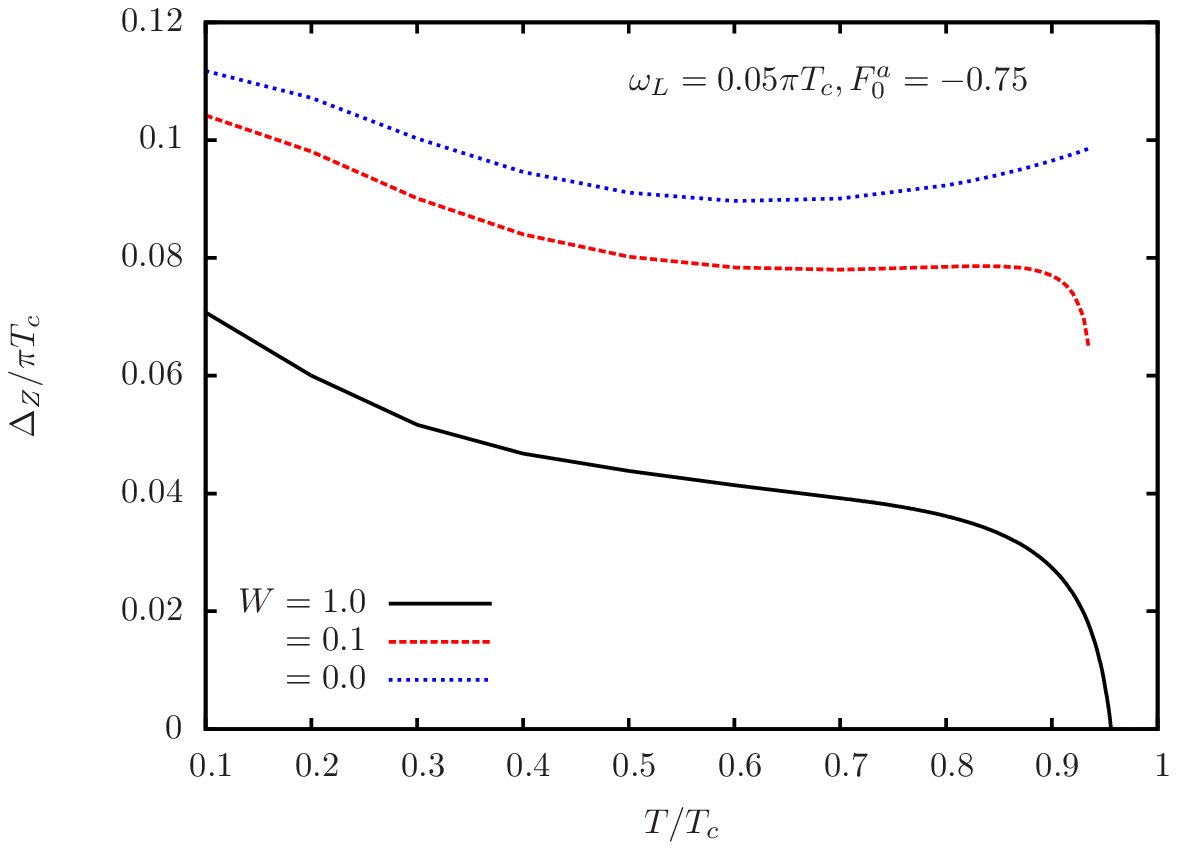}
 \includegraphics[width=0.49\columnwidth]{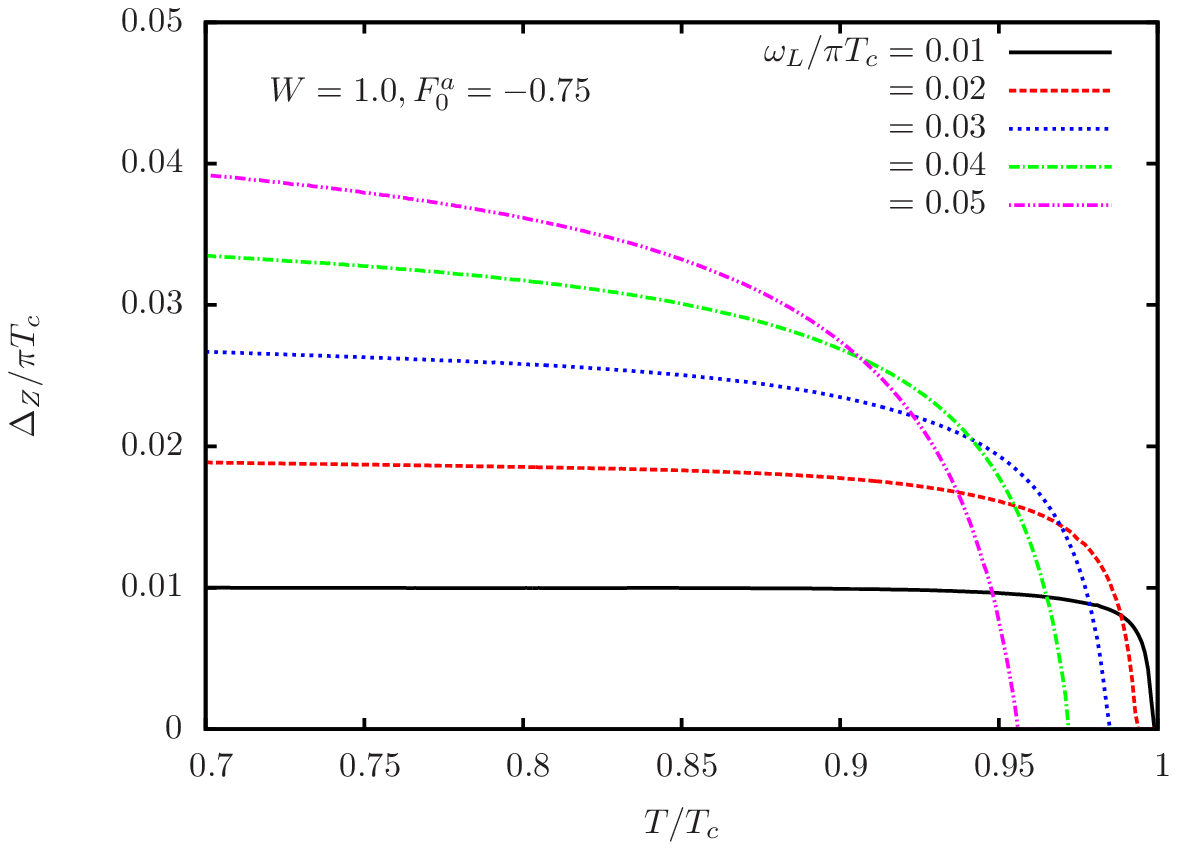}
\end{center}
\caption{\label{fig:dzT} Temperature dependence of ``Zeeman gap''
 $\Delta_{Z}$.}
\end{figure}

``Zeeman gap'' $\Delta_Z$ is the lower edge of the surface bound state 
band.
Temperature dependence of $\Delta_Z$ calculated 
numerically using the self-consistent  order parameter is shown in
Fig.\ref{fig:dzT}. 
$\Delta_Z$ for the diffusive surface $W\neq0 $
is lowered than that for the specular surface by the level repulsion
with the continuum states
as discussed in the previous section. 
In the  diffusive limit $W=1$, $\Delta_{Z}$ 
decreases with the temperature and becomes continuously to 
zero at the A-B transition temperature $T_{\rm AB}$ as is shown in the
right panel of Fig.~\ref{fig:dzT}.

\section{Transverse Acoustic Impedance}\label{sec:6}
\def\cG{\check{G}}
Acoustic impedance is  a useful probe to study the surface as well
as
the bulk properties of neutral liquid ${}^3$He.\cite{Halperin} 
In this section, we consider the transverse acoustic impedance of 
superfluid ${}^3$He-B under the magnetic field normal to the oscillating 
surface.

We consider the rough wall
oscillating in $x$-direction like $R(t)=R e^{-i\Omega t}$.
The acoustic impedance $Z$ is defined by a ratio of the stress
tensor $\Pi_{xz}$ of the liquid at the wall to the velocity $\dot{R}$
of the wall
\begin{equation}
 Z\equiv Z'+i Z''=\frac{\Pi_{xz}}{\dot{R(t)}}
\end{equation}
To study the time dependent problem we use the Keldysh Green function
formalism. We have already reported a theory for the system without
magnetic field.\cite{Nagai2008,Nagato2007} 
In case  of the magnetic field normal to
the surface, we can formulate in  almost the same way as the previous
report. Some details are discussed in Appendix \ref{app3}. 
The final expression
for the transverse acoustic impedance is given in a form
\begin{align}
\frac{Z}{Z_N}
&=\int\frac{d\epsilon}{\Omega}
\bigg( h_-F(\epsilon_+,\epsilon_-)^R-h_+F(\epsilon_+,\epsilon_-)^A
 +(h_+-h_-)F(\epsilon_+,\epsilon_-)^a\bigg),\label{z_of_B}
\end{align}
where $\epsilon_\pm=\epsilon\pm \Omega/2$, $h_{\pm}=\tanh(\epsilon_{\pm}/2T)$, $Z_N$ is the normal state
impedance in the diffusive limit ($W=1$).
The retarded function $F^R$, the advanced function $F^A$ and the
anomalous
function $F^a$
are defined by
\begin{equation}
 F^R=F(\epsilon_+ +i0,\epsilon_- +i0),\quad
 F^A=F(\epsilon_+ -i0,\epsilon_- -i0),\quad
 F^a=F(\epsilon_+ +i0,\epsilon_- -i0).
\end{equation} 
\noindent
Explicit form of $F$ 
is given in Appendix~\ref{app3}.

\begin{figure}
\begin{center}
 \includegraphics{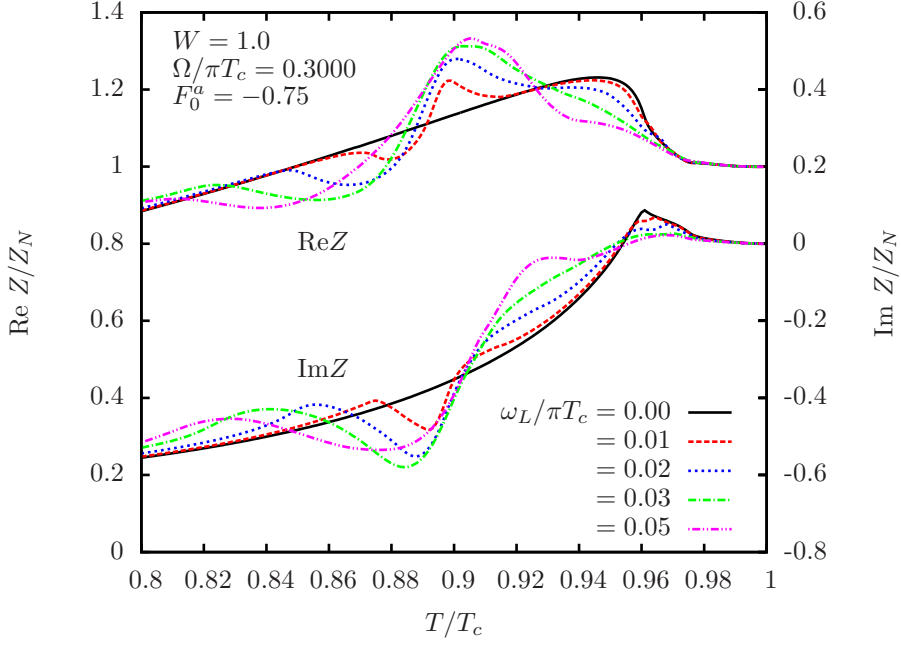}
\end{center}
 \caption{\label{fig:Z_Om} Temperature dependence of
 the acoustic impedance $Z$ for a fixed frequency $\Omega=0.3\pi T_c$. }
\end{figure}

We numerically calculate the acoustic impedance $Z=Z'+i Z''$ using
the self-consistent order parameters. 
In Fig. \ref{fig:Z_Om}, we show the temperature dependence of the
acoustic impedance for a fixed frequency $\Omega=0.3\pi T_c$ under
magnetic fields $\omega_L/\pi T_c=0.0,0.01,0.02,0.03$ and $0.05$. 
In the absence of magnetic field, $Z$ is characterized by a kink in 
the real part
$Z'$ and a peak in the imaginary part $Z''$ at the temperature
at which the condition $\Omega=\Delta^*(T)+\Delta_{\rm bulk}(T)$ is
satisfied. It has been shown that the kink-and-peak structure comes 
from the pair excitations of the bound state and the Bogoliubov
quasi-particle.\cite{Nagai2008, Aoki, Nagato2007} 
Under finite magnetic fields, the overall temperature dependence of
$Z'$ and $Z''$ are shifted to lower temperature. 
The peak in the imaginary part $Z''$ 
found under zero magnetic field eventually disappears.
These behaviors are in agreement with a recent experimental
report by Akiyama et al.\cite{Akiyama2015}

\begin{figure}
\begin{center}
 \includegraphics{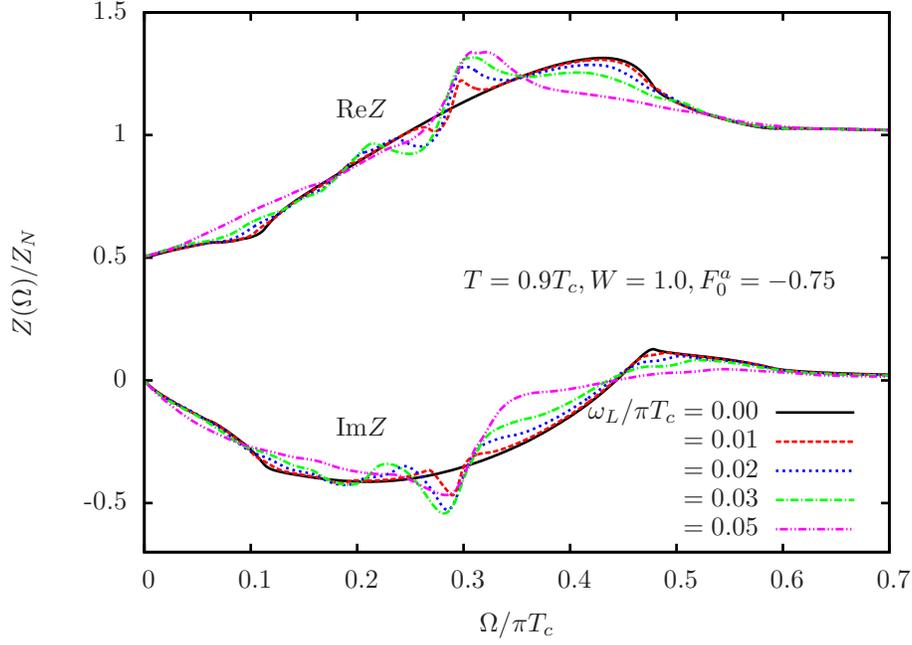}
\end{center}
 \caption{\label{fig:Z_W} Frequency dependence of acoustic impedance
 $Z$ at $T=0.9T_c$}
\end{figure}

At higher magnetic field, the second peak in $Z'$ appears at lower
temperature and the impedance shows bumpy structure.
The similar bumpy structure can be seen also in the frequency dependence
at fixed temperatures.
In Fig. \ref{fig:Z_W}, we show typical results of the frequency 
dependence of $Z(\Omega)$ at $T=0.9T_c$ under various magnetic
fields. One can find a large bump around $\Omega=\Delta_{\rm bulk}\sim
0.3\pi T_c$ when $\omega_L\neq0$. At lower temperatures, the bumpy
structure becomes not so pronounced.

\begin{figure}
\begin{center}
 \includegraphics{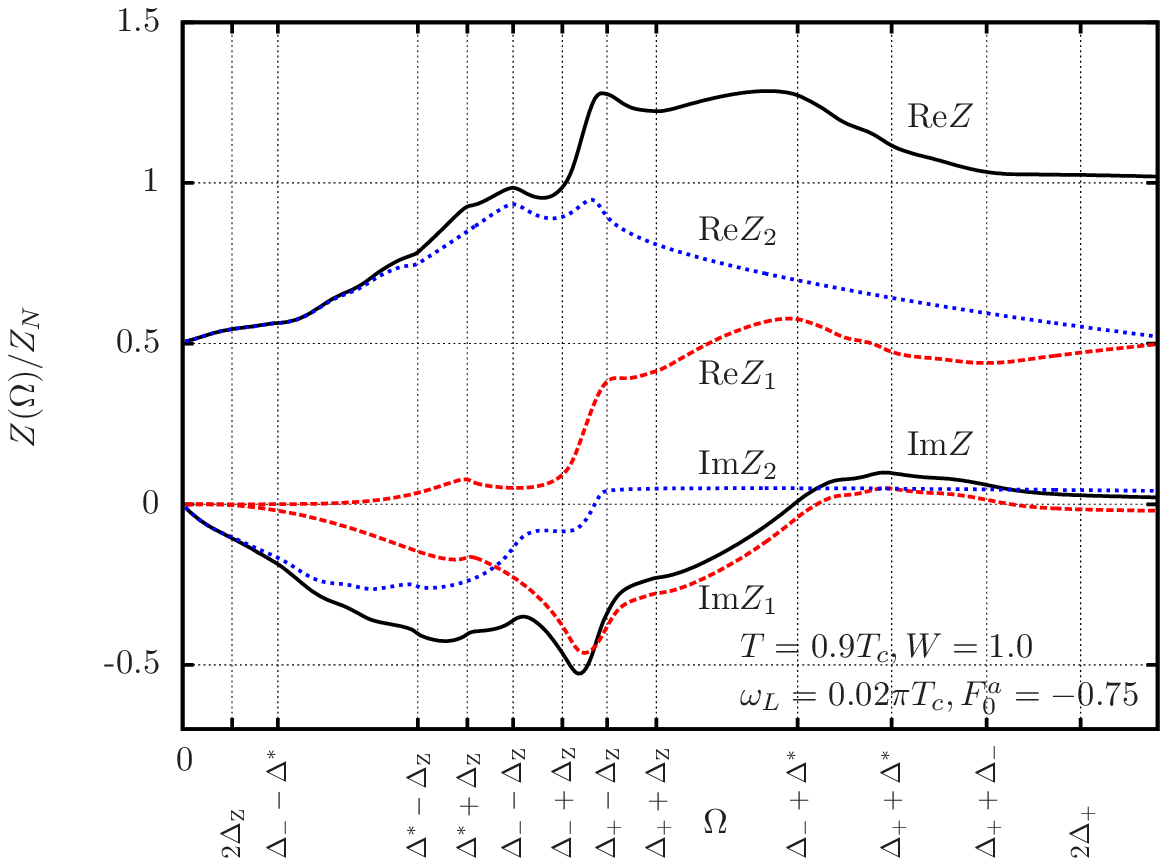}
\end{center}
\caption{\label{fig:Z_Wseparated} Separated transverse acoustic 
 impedances.
 The dashed and dotted lines denote the separated $Z_1$ and $Z_2$
 components. 
 Upper three curves are real parts  and lower three curves are imaginary parts.
}
\end{figure}

To study the origin of the bumpy structure, 
we divide the acoustic impedance $Z$ into two parts 
$Z=Z_1+Z_2$ 
by separating the
energy integral region in Eq.~(\ref{z_of_B}) into $|\epsilon|<\Omega/2$
for $Z_1$ and $|\epsilon|>\Omega/2$ for $Z_2$ (see  
Ref.~\onlinecite{Nagai2008}).
The component $Z_1$ and $Z_2$ give
the contribution from the pair excitations and 
the contribution from the scattering of thermally occupied quasi-particle
states, respectively. This can be understood when we look into the
real part of $Z$ that is related to the energy absorption.
\begin{align}
 Z_1'/Z_n&=\int_{-\Omega/2}^{\Omega/2}\,\frac{d\epsilon}{\Omega}
2[f(\epsilon_+)+f(|\epsilon_-|)-1]\left(\frac{F^R+F^A}{2}-\Re F^a\right)\\
 Z_2'/Z_n&=2\int_{\infty}^{\Omega/2}\,\frac{d\epsilon}{\Omega}
2[f(\epsilon_+)-f(|\epsilon_-|)]\left(\frac{F^R+F^A}{2}-\Re F^a\right)
\end{align}
where $f$ is the Fermi distribution function. Combination of the
Fermi distribution functions implies that $Z_1$ is from the
pair excitations and $Z_2$ is from the scattering of thermally occupied quasi-particle
states. 

In Fig. \ref{fig:Z_Wseparated} we show the frequency dependence of 
the separated acoustic impedances under the magnetic field at $T=0.9T_c$
and $W=1$. 
We find that the bump in $Z'$ is formed by a peak in $Z'_2$ around
$\Omega=\Delta_--\Delta_Z$ 
and a rapid increase of $Z'_1$ in the range 
$\Delta_-+\Delta_Z < \Omega < \Delta_+-\Delta_Z$.
One can also find well defined peak at $\Delta_- -\Delta_Z$ in $Z_2'$
and at $\Delta_Z+\Delta^*$ in $Z_1'$. Precise measurements will provide
useful informations for characteristic energies
$\Delta_Z$, $\Delta^*$, $\Delta_-$ and $\Delta_+$. 
One may expect that there exists a jump at $\Omega=2\Delta_Z$ in the
pair excitation spectrum $Z_1'$. Unfortunately this is not true
because the frequency window $\Delta_Z < \epsilon < \Omega/2$ for $Z_1$ becomes
small at $\Omega=2\Delta_Z$.

\begin{figure}
\begin{center}
 \includegraphics{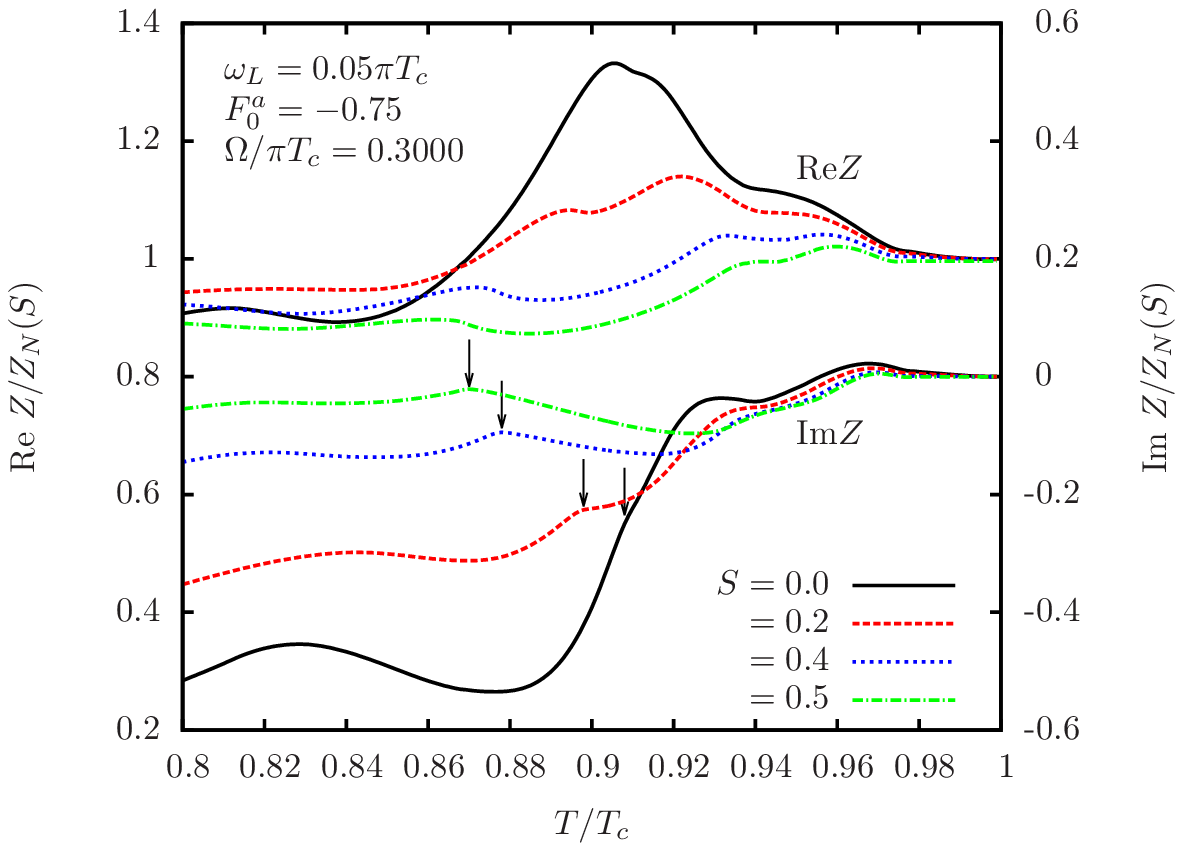}
\end{center}
\caption{\label{fig:Z_T} Temperature dependence of the acoustic
 impedance $Z$ for several values of specularity $S$. 
$Z$ is scaled
 by the normal state value of the same specularity,
 $Z_N(S)=Z_N(0)(1-S)$. 
Vertical arrow indicates the
temperature at which $\Omega=\Delta_+-\Delta_Z$ for each $S$.
}
\end{figure}

In liquid $^3$He, the surface boundary condition
for the quasi-particles can be changed by coating the wall by $^4$He
atoms. It has been suggested that the $^4$He film 
becomes a superfluid and the specularity 
is enhanced.\cite{Okuda, Murakawa, Murakawa2011, Freeman,Tholen,Kim, MurakawaKT}
In Fig. \ref{fig:Z_T}, we show the temperature dependence of $Z$
under a fixed frequency 
for several values of  surface specurality $S$ 
which is related to the roughness parameter $W$.\cite{Nagai2008,roughp}
We have chosen $S=0.0, 0.2, 0.4$ and $0.5$, which correspond to 
$W=1.0,0.264,0.138$ and $0.1$, respectively.
When $S < 0.50$, $W$ is rather small. It follows that
the subgap disappears and dominant characteristic energies
which remain are $\Delta_Z$ and $\Delta_+$.
As indicated by vertical arrows in Fig.~\ref{fig:Z_T},
we find a small peak in the imaginary part $Z$'' at 
the temperature $T(S,\Omega)$ at which $\Omega=\Delta_+(T)-\Delta_Z(T)$
is satisfied. $T(S,\Omega)$ decreases as $S$ increases. This is explained
from the fact that
$\Delta_+-\Delta_Z$ is a decreasing function of temperature
and also is an increasing function of $S$ because $\Delta_Z$ increases
with $S$ while $\Delta_+$ is independent of $S$.    

\section{Summary and Discussion}
We have shown that the subgap structure is a result of the
interplay between the surface roughness and the surface Majorana
state. Subgap structure is also reported in two-dimensional
superconductors.\cite{subgap, Golubov} Study of the
subgap structure in superfluid $^3$He-B and $p$-wave superconductors
will contribute to the further understanding of the surface Majorana
Fermions.

In this paper, the surface states of superfluid $^3$He-B 
with a rough plane surface is discussed. We considered the effect
by magnetic field applied perpendicular to the surface. 
We calculated the self-consistent order parameter, spin polarization
density and the surface density of states using the quasi-classical
Green function theory with the random $S$-matrix model for the surface
roughness.

It was shown that the subgap exists also in $^3$He-B under  finite 
magnetic field. The surface density of states is 
characterized by 
the ``Zeeman gap'' $\Delta_Z$, 
the upper edge of the surface bound states $\Delta^*$,
and the Zeeman splitted bulk energy gap $\Delta_\pm$.
We discussed the dependence of such characteristic energies on
the magnetic field, the temperature, and the surface roughness
parameter $W$.

We have also reported the
results of numerical calculations of the transverse acoustic impedance
and discussed its magnetic field dependence. We found bumpy behavior
in the frequency as well as the temperature dependence of the
acoustic impedance and discussed how this behavior is related to
the surface density of states.
 
It is a future problem  how to detect the ``Zeeman gap'' $\Delta_Z$
experimentally.
Recently, using micro-electro-mechanical system (MEMS) devices 
in $^3$He-B, Zheng et al.\cite{Zheng,Zheng2017} reported an anomalous damping 
thought to be caused by the surface bound states.
It is of great interest how the damping is  influenced by magnetic field.

In this paper we considered only the magnetic field 
perpendicular to the surface. The effect by the magnetic field
parallel to the surface is an interesting  problem. Quasi-classical
theory of the system with magnetic field parallel to the surface
has not been fully developed.
This will be discussed in a forthcoming paper.

\begin{acknowledgments}
We thank R. Nomura and Y. Okuda for discussion and showing their experimental
results before publication. 
This work was supported in part by JSPS KAKENHI Grant Number JP26400364. 
\end{acknowledgments}

%

\appendix
\section{ $\hat{G}_s$ for constant order parameter}\label{app1}
In this appendix, we show the explicit form of $\hat{G}_s$ at the surface
when the order parameter $\Delta_\parallel, \Delta_\perp$ are constant.
Substituting Eq.(\ref{bulk}) into Eq. (\ref{specularG}), we find $\hat{G}_s$
\begin{align}
\hat{G}_{s11}&=-\hat{G}_{s44}=\frac{i}{2}
\left(
\dfrac{E_+
(1-\epsilon/\sqrt{\epsilon^2-\Delta_t^2})}
{h-\sqrt{\epsilon^2-\Delta_t^2}}
+\dfrac{E_-
(1+\epsilon/\sqrt{\epsilon^2-\Delta_t^2})}
{h+\sqrt{\epsilon^2-\Delta_t^2}}
\right) \\
\hat{G}_{s22}&=-\hat{G}_{s33}=-\frac{i}{2}\left(\dfrac{E_+(1+\epsilon/\sqrt{\epsilon^2-\Delta_t^2})}{h-\sqrt{\epsilon^2-\Delta_t^2}}
+\dfrac{E_-(1-\epsilon/\sqrt{\epsilon^2-\Delta_t^2})}{h+\sqrt{\epsilon^2-\Delta_t^2}}
\right) \\
\hat{G}_{s12}&=\hat{G}_{s21}=-\hat{G}_{s34}=-\hat{G}_{s43}=-\frac{\Delta_t\Delta_\ell}{\epsilon^2-(h^2+\Delta_t^2)}\\
\hat{G}_{s13}&=\hat{G}_{s31}=\hat{G}_{s24}=\hat{G}_{s42}=
\frac{\Delta_t}{2}\left(\frac{E_+/\sqrt{\epsilon^2-\Delta_t^2}}
{h-\sqrt{\epsilon^2-\Delta_t^2}}
-\frac{E_-/\sqrt{\epsilon^2-\Delta_t^2}}
{h+\sqrt{\epsilon^2-\Delta_t^2}}\right)\\
\hat{G}_{s14}&=\hat{G}_{s41}=i\frac{\Delta_\ell(h-\epsilon)}{\epsilon^2-(h^2+\Delta_t^2)}\\
\hat{G}_{s23}&=\hat{G}_{s32}=-i\frac{\Delta_\ell(h+\epsilon)}{\epsilon^2-(h^2+\Delta_t^2)},
\end{align}
where $E_\pm$ is given by Eq.(\ref{q_particle}).
It is obvious  that $\hat{G}_s$ satisfies Eq.(\ref{symmetry})
\begin{equation}
 \rho_2\sigma_1 \hat{G}_s \sigma_1 \rho_2= -\hat{G}_s
\end{equation}
and also $\hat{G}_s$ is a symmetric matrix
\begin{equation}
 \hat{G}_s={}^t\hat{G}_s.
\end{equation}
\section{Riccati representation of Quasi-classical Green
 function}\label{app2}
In this appendix, we consider the Riccati
representation\cite{rough,roughp,schopohl,eschrig}
 of 
the quasi-classical Green function $\hat{G}_\alpha$ 
and the rough surface boundary condition.

Here we consider $\hat{G}_\alpha$ for $\phi=\pi/2\ (K_x=0)$ as in 
section~\ref{section:3}. 
The quasi-classical Green's function $\hat G_+$ in the present geometry
can be parametrized
in terms of  $2\times2$ spin space matrices $\cal D_+$ and 
$\mathcal{F}_+$ as\cite{eschrig}
\begin{align}
 \hat{G}_+(K,z,\epsilon)&=i
\begin{pmatrix}
1 & -i \mathcal{F}_+ \\
i \mathcal{D}_+ & 1
\end{pmatrix}
\begin{pmatrix}
1 &  \\
 & -1
\end{pmatrix}
\begin{pmatrix}
1 & -i \mathcal{F}_+ \\
i \mathcal{D}_+ & 1
\end{pmatrix}^{-1}.\label{gp}
\end{align}
From the relation $\hat{G}_-(z)=-\rho_2\sigma_1\hat{G}_+(z)\sigma_1\rho_2$ of
Eq. (\ref{g:symmetry}), we find
\begin{align}
 \hat{G}_-(K,z,\epsilon)&=i
\begin{pmatrix}
1 & -i \sigma_1\mathcal{D}_+\sigma_1 \\
i \sigma_1\mathcal{F}_+ \sigma_1& 1
\end{pmatrix}
\begin{pmatrix}
1 &  \\
 & -1
\end{pmatrix}
\begin{pmatrix}
1 & -i \sigma_1\mathcal{D}_+\sigma_1 \\
i \sigma_1\mathcal{F}_+\sigma_1 & 1
\end{pmatrix}^{-1}.\label{gm}
\end{align}
The matrices $\cal D_+$ and $\mathcal{F}_+$ obey the following Riccati type
differential equations 
\begin{align}
 - iv_F \cos\theta\,\partial_z {\cal D}_+(z) =&
 \begin{pmatrix}
  1 & i{\cal D}_+(z)
 \end{pmatrix}\rho_2 {\cal E}_+
 \begin{pmatrix}
  1 \cr i{\cal D}_+(z)
 \end{pmatrix}~\label{dr}\\
 iv_F \cos\theta\,\partial_z \mathcal{F}_+(z) =&
 \begin{pmatrix}
  \mathcal{F}_+(z) & i
 \end{pmatrix}\rho_2 {\cal E}_+
 \begin{pmatrix}
  \mathcal{F}_+(z)\cr i
 \end{pmatrix},~
\end{align}
where $\mathcal{E}_\alpha$ is given by Eq. (\ref{EE})
\begin{align}
 {\cal E}_\alpha &= \begin{pmatrix}
			    \displaystyle
			    \epsilon + \tilde{h}\sigma_3 &
			    \Delta_\alpha \cr
			    \Delta_{-\alpha} &
			    \displaystyle
			    -\epsilon + \tilde{h}\sigma_3\cr
 \end{pmatrix}.
\end{align}
When $\mathrm{Im}\, \epsilon > 0$, one should solve the Riccati equation for
$\mathcal{D}_+$ in the direction from $z=\infty$ to $z=0$ and for
$\mathcal{F}_+$ in the opposite direction to have  stable solutions. The
boundary condition for $\mathcal{D}_+$  is given at $z=\infty$ by
\begin{align}
 i\mathcal{D}_+(\infty)&=
 (\hat{G}^{(0)}_+)_{21}\frac{1}{(\hat{G}^{(0)}_+)_{11}+i}
\nonumber\\
&= \frac{i}{\alpha_+\beta_- -\alpha_-\beta_+}
\begin{pmatrix}
\beta_- -\beta_+ & \alpha_+ -\alpha_- \cr
(\alpha_+-\alpha_-)\beta_+\beta_- & \beta_- -\beta_+
\end{pmatrix},\label{dinfty}
\end{align}
where $\hat{G}^{(0)}$ is the bulk Green's function given by
Eq. (\ref{bulk})
and 
\begin{equation}
\alpha_\pm=\frac{\epsilon\pm\sqrt{\epsilon^2-\Delta_t^2}}{\Delta_t},
\qquad
\beta_\pm=i\frac{\tilde{h}\pm \sqrt{\epsilon^2-\Delta_t^2} -E_\mp}{\Delta_l}.
\nonumber
\end{equation}
From Eqs. (\ref{dr}) and (\ref{dinfty}), we find that
\begin{equation}
 \sigma_1 \mathcal{D}_+(z) \sigma_1={}^t\mathcal{D}_+(z).\label{dtranspose}
\end{equation}

The boundary condition for $\mathcal{F}_+$ should be given st $z=0$.
According to the random S-matrix model\cite{rough,roughp, nagai_Verditz} for the rough
surface, $\hat{G}_+$ and $\hat{G}_-$ satisfy at the surface
\begin{align}
 \hat{G}_-(0)=T \hat{G}_+(0) T^{-1}, \qquad T=\frac{1-i\Sigma}{1+i\Sigma},
\label{fboundary}
\end{align}
where $\Sigma$ is the surface self energy given by Eq.(\ref{self}).
It follows from Eqs. (\ref{gp}), (\ref{gm}) and (\ref{fboundary})
that at $z=0$
\begin{equation}
\begin{pmatrix}
1 & -i \sigma_1\mathcal{D}_+\sigma_1 \\
i \sigma_1\mathcal{F}_+ \sigma_1& 1
\end{pmatrix}
= T 
\begin{pmatrix}
1 & -i \mathcal{F}_+ \\
i \mathcal{D}_+ & 1
\end{pmatrix}
\begin{pmatrix}
A & \cr
 & B  
\end{pmatrix},
\end{equation}
where $A, B$ are constant matrices.
The second column reads
\begin{align}
 -i\,^t\mathcal{D}_+&=\left(
T_{11}(-i\mathcal{F}_+)+T_{12}\right)B \\
  1&=\left(T_{21}(-i\mathcal{F}_+)+T_{22} 
\right)B
\end{align}
from which we obtain
\begin{equation}
 i\mathcal{F}_+(0)=\frac{1}{T_{11}+i\,^t\mathcal{D}_+(0) T_{21}}
(T_{12}+i\,^t\mathcal{D}_+(0) T_{22})
\label{ans1}
\end{equation}
From the first column, we can obtain an equivalent result by using
Eq.~(\ref{dtranspose}) and noting from Eqs.~(\ref{symmetry}) and
(\ref{self}) 
that the 
surface self energy $\Sigma$ can be parametrized in a form
\begin{equation}
  \Sigma=
\begin{pmatrix}
s_{11}& 0     & 0     & s_{14}\\
0     & s_{22}& s_{23}& 0 \\
0     & s_{23}& -s_{22} & 0 \\
s_{14}& 0     &     0   & -s_{11}
\end{pmatrix}.\label{sigma:form}
\end{equation}
We can also show that
\begin{equation}
 {}^t\mathcal{F}_+(z)=\sigma_1 \mathcal{F}_+(z) \sigma_1.
\end{equation}

Finally,  ${\hat G}_s$ for the specular surface is given by 
\begin{align}
 \hat{G}_s(K,0,\epsilon)&=i
\begin{pmatrix}
1 & -i {}^t\mathcal{D}_+(0) \\
i \mathcal{D}_+(0) & 1
\end{pmatrix}
\begin{pmatrix}
1 &  \\
 & -1
\end{pmatrix}
\begin{pmatrix}
1 & -i {}^t\mathcal{D}_+(0) \\
i \mathcal{D}_+(0) & 1
\end{pmatrix}^{-1}.
\end{align}
\section{Transverse Acoustic Impedance}\label{app3}
We consider the rough wall
oscillating in $x$-direction like $R(t)=R e^{-i\Omega t}$.
The acoustic impedance $Z$ is defined by a ratio of the stress
tensor $\Pi_{xz}$ of the liquid at the wall to the velocity $\dot{R}$
of the wall
\begin{equation}
 Z=\frac{\Pi_{xz}}{\dot{R(t)}}
\end{equation}

To discuss the time dependent problem 
we use the Keldysh
quasi-classical Green's function
\begin{equation}
 \cG_{\alpha}(\bm{K},z, t, t')=
\begin{pmatrix}
\hat{G}_{\alpha}^R(\bm{K},z, t, t') & \hat{G}_{\alpha}^K(\bm{K},z, t, t')\cr
 & \hat{G}_{\alpha}^A(\bm{K},z, t, t')
\end{pmatrix}
\end{equation}
We can apply the random S-matrix model for the rough surface to the
Keldysh Green function 
in the same manner as to the equilibrium Green
function.\cite{Nagai2008,Nagato2007}
To calculate the stress tensor $\Pi_{xz}$,
we treat the wall displacement $R(t)=Re^{-i\Omega t}$ by perturbation theory.
The stress tensor can be calculated from the Keldysh part of the
equal time Green function
\begin{eqnarray}
 \Pi_{xz}&=&\sum_{K,\alpha}\frac{1}{2v_K}\frac{1}{2}\mathrm{Tr}
\left[K_x\,\alpha
 v_K\frac{i}{2}\delta\hat{G}^K_{\alpha\alpha}(K,0,t,t)\right],
\label{stress_tensor}
\end{eqnarray}
where $v_K=v_F\cos\theta_K$ is the $z$ component of the Fermi velocity.
Following the prescription developed in Ref.~\onlinecite{Nagato2007}, 
we obtain
\begin{eqnarray}
 \Pi_{xz}=\frac{-1}{4}\sum_K\int\frac{d\epsilon}{2\pi}
\mathrm{Tr}\left[ \check{\mathcal{G}}(\epsilon_+)\delta\check{\Sigma}(\epsilon_+,\epsilon_-)
\check{\mathcal{G}}(\epsilon_-)\cG_s(\epsilon_-)
-
\cG_s(\epsilon_+)\check{\mathcal{G}}(\epsilon_+)\delta\check{\Sigma}(\epsilon_+,\epsilon_-)
\check{\mathcal{G}}(\epsilon_-)
\right]^K,\label{stress0}
\end{eqnarray}
where $\epsilon_\pm=\epsilon\pm\Omega/2$,
$\check{\mathcal{G}}(\epsilon), \cG_s(\epsilon)$ are the Fourier
transform of the Keldysh Green functions
$\check{\mathcal{G}}(t-t'), \cG_s(t-t')$ in the equilibrium state
and
\begin{eqnarray}
 \delta\check{\Sigma}(\epsilon,\epsilon')&=&2\pi\delta(\epsilon-\epsilon'-\Omega)
\left(\delta\check{\Sigma}^{\mathrm{D}}+\delta\check{\Sigma}^{\mathrm{OD}}\right),\label{O-D}\\
\delta\check{\Sigma}^{\mathrm{D}}&=& -iK_x R
\left(\check{\Sigma}(\epsilon')-\check{\Sigma}(\epsilon)\right),\label{self1}\\
 \delta\check{\Sigma}^{\mathrm{OD}}&=&
2iW<Q_xR\check{\mathcal{G}}(\epsilon)\left(\cG_s(\epsilon')-\cG_s(\epsilon)\right)
\check{\mathcal{G}}(\epsilon')>_Q\nonumber\\
& &+2W<\check{\mathcal{G}}(\epsilon)\delta\check{\Sigma}^{\mathrm{OD}}
\check{\mathcal{G}}(\epsilon')>_Q.\label{self2}
\end{eqnarray}
The superscript D (OD) means that the
contribution to the impedance comes out through the coupling
with the diagonal (off-diagonal) element of $\check{\mathcal{G}}$ in 
particle-hole space.

Introducing a unitary matrix 
\begin{align}
   {\hat\gamma} = \begin{pmatrix}
	     1 & 0 & 0 & 0 \cr
	     0 & 0 & 0 & 1 \cr
	     0 & 0 & 1 & 0 \cr
	     0 & -1 & 0 & 0 \cr
	    \end{pmatrix}~,
\end{align}
we can transform $\hat G_s$, $\Sigma$ and $\cal G$ into the form
\begin{align}
\Sigma &= {\hat\gamma}^\dagger\begin{pmatrix}s_1&0\cr0&-s_2\end{pmatrix}{\hat\gamma},\quad
{\hat G}_s ={\hat\gamma}^\dagger
 \begin{pmatrix}g_1& g_o e^{-i\phi'}\cr g_o e^{i\phi'}&-g_2\end{pmatrix}{\hat\gamma}\\
 {\cal G}&=\left[{\hat G}_s^{-1}-\Sigma\right]^{-1}={\hat\gamma}^\dagger\begin{pmatrix}{\cal G}_1& {\cal G}_o e^{-i\phi'}\cr {\cal G}_o
 e^{i\phi'}&-{\cal G}_2\end{pmatrix}{\hat\gamma}
\end{align}
where $\phi'=\phi_K-\pi/2$. This can be shown
from the properties of $\hat{G}_s$ given by Eqs.~(\ref{rphi}) and 
(\ref{g:symmetry}) and  also from Eq.~(\ref{sigma:form}) for $\Sigma$.
Substituting them 
into Eqs. (\ref{stress0}), we can 
analytically perform the $\phi_K$ average.  We 
finally obtain the expression for the transverse acoustic impedance.
\begin{align}
 Z/Z_N= 
 \int_{-\infty}^{\infty}{d\epsilon}\frac1\Omega \left[
 F^R(\epsilon_+,\epsilon_-) h_-
 - F^A(\epsilon_+,\epsilon_-) h_+
 + F^a(\epsilon_+,\epsilon_-)
 \left(h_+-h_-\right)
 \right]
\end{align}
where $Z_N$ is the normal state impedance and $h_\pm=\tanh\left(\beta\epsilon_\pm/2\right)$. 
The retarded, advanced and anomalous functions $F^{R,A,a}$
are given by
\begin{equation}
 F^R=F(\epsilon_+ +i0,\epsilon_- +i0),\quad
 F^A=F(\epsilon_+ -i0,\epsilon_- -i0),\quad
 F^a=F(\epsilon_+ +i0,\epsilon_- -i0).
\end{equation} 
The explicit form of $F$ is given as follows.
\begin{align}
 F(\epsilon_+,\epsilon_-)&=F^D(\epsilon_+,\epsilon_-)+F^{OD}(\epsilon_+,\epsilon_-)\\
 F^D&=\left\langle \sin^2\theta ~\frac14{\rm Tr}
 \left[ (s_1'-s_1)a_1 + (s_2'-s_2)a_2 \right]\right\rangle_\theta\\
 F^{OD}&=\left\langle \sin\theta ~\frac14{\rm Tr}
 \left[ \bar\zeta ~b_1 - \bar\eta ~b_2 \right]\right\rangle_\theta\\
 a_1 &= 
 {\cal G}_1(g_1'-g_1){\cal G}_1' + {\cal G}_1(g_o'-g_o){\cal G}_o'
+ {\cal G}_o(g_o'-g_o){\cal G}_1' - {\cal G}_o(g_2'-g_2){\cal G}_o' \\
 a_2 &=
 {\cal G}_2(g_2'-g_2){\cal G}_2' + {\cal G}_2(g_o'-g_o){\cal G}_o'
+ {\cal G}_o(g_o'-g_o){\cal G}_2' - {\cal G}_o(g_1'-g_1){\cal G}_o' \\
 b_1 &= 
 {\cal G}_1(g_1'-g_1){\cal G}_o' + {\cal G}_o(g_o'-g_o){\cal G}_o'
- {\cal G}_1(g_o'-g_o){\cal G}_2' + {\cal G}_o(g_2'-g_2){\cal G}_2' \\
 b_2 &=
 {\cal G}_o(g_1'-g_1){\cal G}_1' + {\cal G}_o(g_o'-g_o){\cal G}_o'
- {\cal G}_2(g_0'-g_0){\cal G}_1' + {\cal G}_2(g_2'-g_2){\cal G}_o' 
\end{align}
Here $s, g, \mathcal{G}$ with prime are functions of $\epsilon_-$
while $s, g, \mathcal{G}$ without prime are functions of $\epsilon_+$.
Corresponding to Eq.~(\ref{self2}) for $\delta\check{\Sigma}^{OD}$
we should solve the following equations for $\bar\zeta$ and $\bar\eta$
\begin{align}
 \bar\zeta &= -2W \langle {\cal G}_1 \bar\zeta {\cal G}_2'\rangle_\theta
 -W\langle \sin\theta ~b_1 \rangle_\theta\\
 \bar\eta &= -2W \langle {\cal G}_2 \bar\eta {\cal G}_1'\rangle_\theta
 +W\langle \sin\theta ~b_2 \rangle_\theta,
\end{align}
where
\begin{align}
\langle\cdots\rangle_\theta=2\int_0^{\pi/2}d\theta \sin\theta\cos\theta \cdots~.
\end{align}


\begin{thebibliography}{99}
\bibitem{AdR} V. Ambegaokar, P. G. de Gennes and D. Rainer,
Phys. Rev. A {\bf 9}, 2676 (1974).
\bibitem{BZ} L. J. Buchholtz and G. Zwicknagl, Phys. Rev. B {\bf 23},
5788 (1981).
\bibitem{HN} J. Hara and K. Nagai, Prog. Theor. Phys. {\bf 76}, 1237
(1986).
\bibitem{Zhang}
W. Zhang, Phys. Lett. A {\bf 130},  314 (1988).
\bibitem{Nagai2008} K. Nagai, Y. Nagato, M. Yamamoto and S. Higashitani,
J. Phys. Soc. Jpn {\bf 77},  111003 (2008).
\bibitem{Schnyder} A. P. Schnyder, S. Ryu, A. Furusaki and A. W. W. Ludwig:
Phys. Rev. B {\bf 78} (2008) 195125.
\bibitem{Qi} X.-L. Qi, T.L. Hughes, S. Raghu and S.-C.Zhang:
Phys. Rev. Lett. {\bf 102} (2009) 187001. 
\bibitem{Okuda} Y. Okuda and R. Nomura, J. Phys.: Condensed Matter {\bf
	24}, 343201 (2012).
\bibitem{Mizushima} T. Mizushima, Y. Tsutsumi, T. Kawakami, M. Sato,
M. Ichioka, and K. Machida, J. Phys. Soc. Jpn {\bf 85}, 022001 (2016).
\bibitem{Aoki}
Y. Aoki, Y. Wada, R. Nomura, Y. Okuda, Y. Nagato,
M. Yamamoto, S. Higashitani and K. Nagai,
Phys. Rev. Lett.  {\bf 95},  075301 (2005).
\bibitem{Murakawa}
S. Murakawa, Y. Tamura, Y. Wada, M. Wasai, M. Saitoh, Y. Aoki,
R. Nomura, Y. Okuda, Y. Nagato,
M. Yamamoto, S. Higashitani and K. Nagai, Phys. Rev. Lett. 
{\bf 103 },  155301 (2009).
\bibitem{Murakawa2011}
S. Murakawa, Y. Wada,  Y. Tamura, M. Wasai, M. Saitoh, Y. Aoki,
R. Nomura, Y. Okuda, Y. Nagato,
M. Yamamoto, S. Higashitani and K. Nagai, J. Phys. Soc. Jpn 
{\bf 80 }, 013602 (2011).
\bibitem{Nagato2007} Y. Nagato, M. Yamamoto, S. Higashitani and K. Nagai,
J. Low Temp. Phys. {\bf 149},  294 (2007).
\bibitem{roughp}
Y. Nagato, M. Yamamoto and K. Nagai,
J. Low Temp. Phys. {\bf 110},  1135 (1998).
\bibitem{Vorontsov}
A.~B. Vorontsov and J.~A. Sauls,
  {Phys. Rev. B} \textbf{{68}}. {064508} ({2003}).
\bibitem{subgap}Y. Nagato, S. Higashitani and K. Nagai,
	J. Phys. Soc. Jpn
 {\bf 80}, 113706 (2011).
\bibitem{Nagato2009} Y. Nagato, S. Higahitani and K. Nagai,
J. Phys. Soc, Jpn, {\bf 78}, 123603 (2009).
\bibitem{Chung} S. B. Chung and S.-C. Zhang, Phys. Rev. Lett. {\bf 103},
235301 (2009). 
\bibitem{Volovik}G. E. Volovik, Pm Zh. Eksp. Teor. Fiz. {\bf 90},
	587, (2009).
\bibitem{Miz} T. Mizushima and K. Machida, J. Low Temp. Phys. {\bf 162},
204 (2011).
\bibitem{Silaev} M. A. Silaev, Phys. Rev. B {\bf 84}, 144508 (2011).
\bibitem{rough} Y. Nagato, S. Higashitani, K. Yamada and K. Nagai,
J. Low Temp. Phys.  {\bf 103},  1 (1996).
\bibitem{nagai_Verditz}
K. Nagai, in {\it Quasiclassical Methods in
	Superconductivity and Superfluidity}, edited by D. Rainer 
and J.A. Sauls (Verditz, Bayreuth, Germany 1996)
\bibitem{aahn} M. Ashida, S. Aoyama, J. Hara and K. Nagai,
Phys. Rev. B {\bf 40},  8673 (1989)
\bibitem{Ovchinnikov} Yu. N. Ovchinnikov, Sov. Phys. JETP {\bf 29}, 853
(1969).
\bibitem{Kopnin} N. B. Kopnin, J. Low Temp.Phys. {\bf 85}, 267 (1991).
\bibitem{Nagato1993} Y. Nagato, K. Nagai and J. Hara,
J. Low Temp. Phys {\bf 93}, 33 (1993)
\bibitem{Higashitani} S. Higashitani and K. Nagai,
J. Phys. Soc. Jpn {\bf 64}, 549 (1995)
\bibitem{schopohl} N. Schopohl and K. Maki, Phys. Rev. B {\bf 52}
490 (1995)
\bibitem{eschrig}
M. Eschrig, Phys. Rev. B {\bf 61} 9061 (2000).
\bibitem{Ashida_Nagai} M. Ashida and K. Nagai, Prog. Theor. Phys.
{\bf 74}, 949 (1985).
\bibitem{Halperin} W.P. Halperin and E. Varoquaux, in {\it Helium
	Three},
ed. by W.P. Halperin and P.L. Pitaevski (Elsevier, Amsterdam,1990), p.353.
\bibitem{Akiyama2015} K. Akiyama, M. Wasai, M. Mashino, T. Nakao,
S. Murakawa, R. Nomura and Y. Okuda, J. Phys. Soc. Jpn {\bf 84},
065001 (2015).
\bibitem{Freeman} M. R. Freeman and R. C. Richardson, Phys. Rev. B
{\bf 41 }, 11011 (1990).
\bibitem{Tholen} S. M. Tholen and J. M. Parpia, Phys. Rev. Lett. {\bf
	68}
2810 (1992).
\bibitem{Kim} D. Kim, M. Nakagawa, O. Ishikawa, T. Hata, T. Kodama
and H. Kojima, Phys. Rev. Lett. {\bf 71}, 1581 (1993).
\bibitem{MurakawaKT} S. Murakawa, M. Wasai, K. Akiyama, Y. Wada,
Y. Tamura, R. Nomura and Y. Okuda, Phys. Rev. Lett. {\bf 108}, 025302 (2012).  
\bibitem{Golubov} S. V. Bakurskiy, A. A. Golubov, M. Yu. Kupriyanov,
K. Yada and Y. Tanaka, Phys. Rev. B {\bf 90}, 064513 (2014).
\bibitem{Zheng} P. Zheng, W. G. Jiang, C. S. Barquist, Y. Lee, and 
H. B. Chan, Phys. Rev. Lett. {\bf 117}, 195301 (2016).
\bibitem{Zheng2017} P. Zheng, W. G. Jiang, C. S. Barquist, Y. Lee, and 
H. B. Chan, Phys. Rev. Lett. {\bf 118}, 065301 (2017).
\end{thebibliography}
\end{document}